\documentclass[10pt,aps,prl,twocolumn,amsmath,floatfix,longbibliography,a4paper]{revtex4-2}

\usepackage{graphicx}
\usepackage{hyperref}
\usepackage{textcomp}
\usepackage{microtype}

\hypersetup{
  unicode=false,            
  pdftoolbar=true,          
  pdfmenubar=true,          
  pdffitwindow=false,       
  pdfstartview={FitH},      
  colorlinks=true,          
  linkcolor=blue,           
  citecolor=red,            
  bookmarksdepth=2,
  pdfauthor={Daria Kyvala (https://orcid.org/0000-0002-2603-6453),
    Jindrich Kolorenc (https://orcid.org/0000-0003-2627-8302)},
  pdftitle={Inelastic electron tunneling through adatoms and molecular
    nanomagnets},
  pdfkeywords={IETS, STS, scanning tunneling spectroscopy, inelastic
    electron tunneling, cotunneling, Hubbard model, adatoms}
}

\let\vec=\mathbf
\graphicspath{{figures/}}
\def\DNc{D_\text{Ni}}
\def\DFe{D_\text{Fe}}

\begin{document}

\title{Inelastic Electron Tunneling through Adatoms and Molecular Nanomagnets}
\author{Daria K\'{y}vala}
\affiliation{Institute of Physics (FZU), Czech Academy of Sciences, Na
  Slovance 2, 182 00 Praha, Czech Republic}
\author{Jind\v{r}ich Koloren\v{c}}
\email{kolorenc@fzu.cz}
\affiliation{Institute of Physics (FZU), Czech Academy of Sciences, Na
  Slovance 2, 182 00 Praha, Czech Republic}

\date{\today}

\begin{abstract}
We discuss a theoretical description of the inelastic electron
tunneling spectra (IETS) of a magnetic nanosystem (an atom or a
molecule) adsorbed on a solid surface measured in a scanning tunneling
microscope (STM). We represent the nanosystem by means of a cluster Hubbard
model, which allows us to study scenarios when the tunneling electrons
sequentially interact with several magnetic centers inside the
nanosystem or when the magnetic centers are made out of heavy atoms
with a strong spin-orbit coupling and large orbital moments. The
sequential tunneling through multiple centers is illustrated
on an adatom probed by an STM tip with a nickelocene molecule attached
to it. For atoms with a large
orbital moment, we find the transitions accessible by IETS to be
governed by the selection rule $\Delta J_z\leq 2\ell+1$, where $J_z$
is the projection of the total angular momentum of the atom to the
quantization axis and $\ell$ is the orbital momentum quantum number of the
partially filled atomic shell carrying the magnetic moment. For atoms
with magnetic moments dominated by spin, the spectra are naturally
dominated by transitions fulfilling the traditional selection rule
$\Delta J_z\leq 1$.


\end{abstract}

\maketitle




The inelastic electron tunneling spectroscopy (IETS) was initially
developed for characterization of molecular vibrations, representing a
complementary tool to the Raman spectroscopy
\cite{jaklevic1966,hansma1977}. The observation
of electronic transitions (including f--f transitions) followed soon
afterwards \cite{leger1972,adane1975,hipps1993}. The measurements were
originally done on
ensembles of molecules placed inside metal-insulator-metal tunnel
junctions but we are
concerned with inelastic spectroscopy of a single atom or molecule in a
scanning tunneling microscope (STM)
\cite{binnig1985,stipe1998}. Namely, we discuss
theoretical description of the spin-excitation spectroscopy
\cite{heinrich2004,hirjibehedin2007} that is usually modeled in
terms of the spin-assisted tunneling Hamiltonians considering an
exchange between the spin of the tunneling electron and
the spin of the nanosystem adsorbed on the surface
\cite{fernandez-rossier2009,fransson2009,persson2009,ternes2015}. We
employ an alternative strategy, working directly with electronic
instead of spin degrees of freedom, which
facilitates a straightforward generalization to scenarios when the
tunneling involves sequential interactions of the tunneling electron
with several magnetic centers (for instance when the STM tip has a
magnetic molecule attached
\cite{ormaza2017,verlhac2019,wackerlin2022}) or when atoms
with large orbital moments are among the magnetic centers in the
nanosystem. On top of that, the electronic model enables a unified
description of IETS and x-ray magnetic circular dichroism
\cite{wackerlin2022}.


\paragraph*{Differential conductance.}
\addcontentsline{toc}{section}{Differential conductance}
We approximate the magnetic nanosystem by the cluster Hubbard
model $\hat{\mathcal H}_{ns}$, each magnetic 
center being represented by one 
site of the model~\cite{chiesa2013}. The metallic electrodes (tip and
surface) are taken as reservoirs of non-interacting electrons
(with Hamiltonians $\hat{\mathcal H}_{t}$ and $\hat{\mathcal
  H}_{s}$). The coupling between the electrodes and the nanosystem is
described by a tunneling operator
\begin{equation}
\hat{\mathcal{V}} = \sum_{ki\Gamma\mu} \Bigl(
A_{ki\Gamma} \hat{a}^\dagger_{ki\Gamma\mu}\hat{d}_{i\Gamma\mu}
+ B_{ki\Gamma} \hat{b}^{\dagger}_{ki\Gamma\mu}  \hat{d}_{i\Gamma\mu} 
+ \text{h.c.}\Bigr),
\label{eq:coupling}	
\end{equation}
where the electron operators $\hat d$ correspond to the
one-electron states defining the magnetic centers, and the
operators $\hat a$ and $\hat b$ correspond to the
conduction-electron states in the tip and surface. Index $k$ is a star 
of a wave vector combined with the band index, $i$ enumerates the
magnetic centers, $\Gamma$ labels irreducible
representations at a given magnetic center, and $\mu$ labels
components of $\Gamma$ if it is multidimensional.
The states $\hat{a}_{ki\Gamma\mu}$ ($\hat{b}_{ki\Gamma\mu}$) are
one-electron eigenstates of
$\hat{\mathcal H}_t$ ($\hat{\mathcal H}_s$) with an energy
$\epsilon_k$. For a magnetic center surrounded by
an isotropic environment, $\mu$ would be $m\sigma$ where $m$ and
$\sigma$ are magnetic and spin quantum numbers, and
$\hat{a}_{ki\Gamma\mu}$ would be spherical waves around the center
$i$. Defined this way, $\hat{\mathcal V}$ conserves
the symmetry of the magnetic centers
\cite{haldane1977,bringer1977,gunnarsson1989}.

The states $\hat{a}_{ki\Gamma\mu}$ and
$\hat{a}_{k'i'\Gamma'\mu'}$ coupled to different centers, $i\not=i'$,
are not mutually orthogonal in general but we impose this orthogonality as an
additional simplification. This simplification does not come into play
when each electrode is coupled to only one magnetic center. In
addition, we
neglect the energy dependence of the tunneling amplitudes $A_{ki\Gamma}$
and $B_{ki\Gamma}$ and replace them with $A_{i\Gamma}$ and
$B_{i\Gamma}$. This is reasonable since magnetic IETS
typically probes the conduction states in the electrodes only in a
small energy window of several tens of meV around the Fermi
level~$\epsilon_\text{F}$, where the energy dependence can be replaced
by a constant.

\begin{figure}
\includegraphics[width=\linewidth]{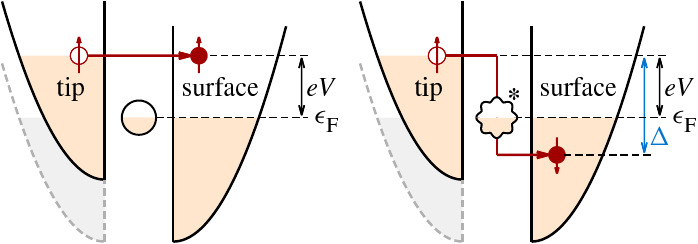}
\caption{Elastic (left) and inelastic
  (right) tunneling. The STM arrangement
  consists of a tip (left electrode), a surface (right electrode),
  and a magnetic nanosystem in between. A voltage~$V$ is applied to
  the tip, which lifts its Fermi level up by the energy
  $eV$. The inelastic channel is active only when the
  voltage exceeds the excitation energy~$\Delta$ of the
  nanosystem, $eV>\Delta$, otherwise there is no empty final state
  available.}
\label{fig:tunneling}
\end{figure}

The many-body eigenstates of $\hat{\mathcal H}_{ns}$
are denoted as $|\psi_{N\alpha}\rangle$, where $N$ is the number of
electrons in the nanosystem and $\alpha$ indexes the states
within the fixed-$N$ subspace of the full Hilbert space. The
corresponding energies are $E_{N\alpha}$. To simplify the notation, we
introduce a class of many-body eigenstates
of the whole system
$
| \chi_{N\alpha} \rangle =
| \phi_t \rangle\otimes
| \phi_s \rangle\otimes
| \psi_{N\alpha} \rangle,
$
where $|\phi_s\rangle$ is the ground state of the surface electrode
(all states
with single-particle energies $\epsilon_k<\epsilon_\text{F}$ are
filled) and $|\phi_t\rangle$ is the ground state of the tip (all states
with single-particle energies $\epsilon_k<\epsilon_\text{F}+eV$ are
filled, since the tip is at a voltage $V$ above the surface,
Fig.~\ref{fig:tunneling}). The energy of $|\chi_{N\alpha}\rangle$
is $\mathcal E_{N\alpha} = E_{N\alpha} + E_t + E_s$, where $E_t$ and
$E_s$ are the ground-state energies of the tip and surface.

Starting from the initial state~$|\chi_{N\alpha}\rangle$, the
tunneling from the tip to the surface proceeds along
one of two paths: (a) an electron
jumps from the tip to the nanosystem and
then the same (or another) electron jumps from the nanosystem to the
surface, or (b) an
electron jumps from the nanosystem to the surface and then an electron
from the tip fills the hole created in the nanosystem.
In~the first case, there is a (virtual) intermediate state $|v \rangle$
with $N+1$ electrons in the nanosystem and with energy $\mathcal
E_v$. In the second case, the intermediate state has $N-1$
electrons in the nanosystem. Such coherent two-step
processes are known as cotunneling~\cite{averin1990,glazman2005}.
There exist tunneling paths
involving virtual states with $N\pm n$ electrons in the nanosystem,
where $n>1$, that correspond to higher orders of the expansion in
the tunneling operator~$\hat{\mathcal{V}}$. We assume that the
tunneling amplitudes $A_{ki\Gamma}$ and $B_{ki\Gamma}$ are small
compared to the charging energy of the nanosystem $U\sim
E_{N\pm1,\alpha} - E_{N\alpha}$ (Coulomb blockade regime), and that
the higher-order processes can be neglected.

We consider the tunneling as a series of independent events, which is
a good approximation as long as the tunneling current
is low and the nanosystem has enough time to decay from an excited
final state before the next tunneling event starts. We also
neglect all temperature effects so that the tunneling always starts
from the ground state of the nanosystem and electrodes. Under these
assumptions, the tunneling current can be calculated from the
Kramers--Heisenberg
formula~\cite{kramers1925,sakurai_aqm_KHformula} that is routinely used in
simulations of another type of inelastic spectroscopy, the resonant inelastic
x-ray scattering (RIXS)~\cite{kotani2001,rueff2010}. In the present
context, the formula reads as
\begin{equation}
I_{\pm}\sim\pm\sum_{f_{\pm}}
\Biggl| \sum_{v} 
 \frac{\langle f_{\pm} | \hat{\mathcal{V}} | v \rangle
  \langle v | \hat{\mathcal{V}} |  \chi_{N\alpha} \rangle}
 {\mathcal E_{v} - \mathcal E_{N\alpha}} 
	\Biggr|^2
  \delta\bigl( \mathcal E_{f_{\pm}} - \mathcal E_{N\alpha}\bigr),
\label{eq:KHformula}
\end{equation}
where the positive current $I_+$ comes from the final states
$| f_{+} \rangle = \hat{b}^{\dagger}_{k'i'\Gamma'\mu '}
\hat{a}_{ki\Gamma\mu} | \chi_{N\beta}\rangle$ with
$\epsilon_{k} < \epsilon_\text{F}+eV$ and 
$\epsilon_{k'}>\epsilon_\text{F}$,
and the negative current $I_-$ comes from the final states
$| f_{-} \rangle = \hat{a}^{\dagger}_{ki\Gamma\mu} \hat{b}_{k'i'\Gamma'\mu'}
| \chi_{N\beta} \rangle$ with $\epsilon_{k} >
\epsilon_\text{F}+eV$ and $\epsilon_{k'}<\epsilon_\text{F}$.
The final-state energies are
$\mathcal E_{f_{\pm}} = {\mathcal E}_{N\beta}
   \mp \epsilon_k \pm \epsilon_{k'}$.
Index~$\alpha$ corresponds to the ground state of
the nanosystem, the final states with $\beta \neq \alpha$ indicate
inelastic channels. If the ground state is degenerate, the
current is computed as an average over the degenerate ground
states.

Fixing the nanosystem at the same voltage as
the surface (Fig.~\ref{fig:tunneling}), which also implies neglecting
any voltage gradient within the nanosystem,  and integrating out the
degrees of freedom of the electrodes, we arrive at a formula for the
differential conductance
\begin{subequations}
\label{eqs:conductance}
\begin{equation}
\frac{\text{d}I}{\text{d}V}=
\sideset{}{_\beta}\sum_{\Delta_{\beta \alpha}<eV}
{\mathcal {G}}_{\alpha\beta}\,
+
\sideset{}{_\beta}\sum_{\Delta_{\beta\alpha}<-eV}
{\mathcal {G}}_{\beta\alpha}\,,
\label{eq:total_conductance}
\end{equation}
where $\Delta_{\beta\alpha}=E_{N\beta}-E_{N\alpha}$ \cite{suppl}. The
first term contributes only at positive biases $V$ (giving
positive current), the second term contributes only at negative
$V$ (giving negative current). The partial conductances
${\mathcal G}_{\alpha\beta}$ have the form
\begin{multline}
\label{eq:partial_conductance}
{\mathcal G}_{\alpha\beta}
=
\frac{2\pi e^2}{\hbar}
\hskip-.7em\sum_{ii'\Gamma\Gamma'\mu\mu'}\hskip-.7em
|A_{i\Gamma}B_{i'\Gamma'}|^2
\rho_{i\Gamma}^t(eV)\rho_{i'\Gamma'}^s(eV-\Delta_{\beta\alpha})
\\
\times
\bigl|\langle \psi_{N\beta}|
\hat{\mathcal O}^{\alpha\beta}_{i\Gamma\mu\,\, i'\Gamma'\mu'}
|\psi_{N\alpha}\rangle\bigr|^2,
\end{multline}
where $\rho_{i\Gamma}^t$ is the density of states (DOS) in the tip projected
onto orbitals $\hat a_{ki\Gamma\mu}$ and
$\rho_{i\Gamma}^s$ is the analogous DOS in the
surface projected onto orbitals $\hat b_{ki\Gamma\mu'}$. Neither
of these DOSes depends on $\mu$ as dictated by symmetry. Further on,
we neglect the energy dependence of the DOSes
analogously as we do for the amplitudes $A_{i\Gamma}$ and $B_{i\Gamma}$.
Finally, the transition operator entering
Eq.~\eqref{eq:partial_conductance},
\begin{multline}
\label{eq:transOp}
\hat{\mathcal O}^{\alpha\beta}_{i\Gamma\mu\,\, i'\Gamma'\mu'}
=
\hat d_{i'\Gamma'\mu'}
\frac1{\hat{\mathcal H}_{ns} - E_{N\alpha} - eV}
\hat d^{\dagger}_{i\Gamma\mu}
\\-
\hat d^{\dagger}_{i\Gamma\mu}
\frac1{\hat{\mathcal H}_{ns} - E_{N\beta} + eV}
\hat d_{i'\Gamma'\mu'}\,,
\end{multline}
\end{subequations}
determines the intensities of the inelastic transitions and their
selection rules. The two terms in
Eq.~\eqref{eq:transOp} are the two cotunneling paths.

The evaluation of the matrix elements of
$\hat{\mathcal O}^{\alpha\beta}_{i\Gamma\mu\,\, i'\Gamma'\mu'}$ is a
many-body problem that we
solve numerically by means of Krylov-subspace methods. The
low-lying eigenstates of the nanosystem $|\psi_{N\beta}\rangle$ are
found using the implicitly restarted Lanczos method \cite{arpack}, the
matrix elements themselves are computed using the band Lanczos 
method~\cite{ruhe1979,meyer1989}.

Our description of IETS is
essentially the same as the low-current and zero-temperature limit of
the theory formulated in~\cite{delgado2011}\footnote{
Eq.~\eqref{eq:KHformula} is the same as the Fermi golden rule
$I_\pm=\pm (2\pi e/\hbar)\sum_{f_{\pm}} |\mathcal H_\text{cotun}|^2 \delta\bigl(
\mathcal E_{f_{\pm}} - \mathcal E_{N\alpha} \bigr)$ 
if the cotunneling Hamiltonian
$\mathcal H_\text{cotun}= \sum_{v}   \protect\langle f_{\pm} | \hat{\mathcal{V}} | v
  \protect\rangle  \protect\langle v | \hat{\mathcal{V}} | \chi_{N\alpha}  \protect\rangle 
  / \bigl(\mathcal E_{v} - \mathcal E_{N\alpha}\bigr)$ is introduced
  as in~\cite{delgado2011}.}. Only the tunneling operator
from~\cite{delgado2011}, with terms like
$\hat a_{k\sigma}^\dagger \hat d_{im\sigma}$ that couple the magnetic
centers to the Bloch waves $\hat a_{k\sigma}$ instead of the
symmetry-adapted waves $\hat a_{ki\Gamma\mu}$, differs from our
tunneling operator, Eq.~\eqref{eq:coupling}. This leads to a
modification of Eq.~\eqref{eq:partial_conductance}; the sums over the
orbital degrees of freedom ($i$ and $m$) are inside the modulus
squared, whereas the sums over the spin $\sigma$ stay outside like in
our result. Such alternative expression for the differential
conductance has a spurious symmetry breaking built in \cite{suppl}.


\paragraph*{The nanosystem.}
\addcontentsline{toc}{section}{The nanosystem}
The applications discussed below fit into a framework of
$\hat{\mathcal H}_{ns} = \sum_{i}\hat{\mathcal H}_{i}
 + \sum_{i<j}\hat{\mathcal{T}}_{ij} + \sum_{i}\hat{\mathcal U}_{i}$,
where $\hat{\mathcal H}_{i}$ characterizes an individual
magnetic center (a partially filled atomic shell with orbital momentum
quantum number~$\ell_i$), $\hat{\mathcal T}_{ij}$ describes
hopping of electrons between two shells, and $\hat{\mathcal U}_{i}$ is the
Coulomb repulsion within a shell. Each shell
\begin{equation}
\label{eq:3d_shell}
\hat{\mathcal H}_{i} = \sum_{\substack{m m'\\ \sigma \sigma'}}
 \Bigl(
  \bigl[
  \zeta^i \vec{l} \cdot \vec{s}
  -\boldsymbol{\mu}\cdot\vec{B}
  \bigr]_{m m'}^{\sigma\sigma'} + 
  h^i_{mm'} \delta_{\sigma \sigma '}
 \Bigr) \hat{d}^{\dagger}_{i m\sigma} \hat{d}_{i m'\sigma'}
\end{equation}
contains the spin-orbit coupling specified by its strength $\zeta^i$,
the interaction of the magnetic
moment $\boldsymbol{\mu}=\mu_\text{B}(\vec{l}+2\vec{s})$ with a
magnetic field $\vec B$, and the crystal-field potential
$h^i$ that we define using the Wybourne parameters $B^i_{kq}$. The
energy of the atomic level $\epsilon^i$ is included
$h^i$ as the $(k,q)=(0,0)$ term with
$B^i_{00}=\epsilon^i$. The hopping operator $\hat{\mathcal{T}}_{ij}$ has
the form
\begin{equation}
\label{eq:T_operator}
\mathcal{\hat{T}}_{ij} = \sum_{\sigma mm'} \Bigl(
 t^{ij}_{mm'} \hat{d}^{\dagger}_{im\sigma} \hat{d}_{jm'\sigma} + 
 \text{h.c.}
 \Bigr)\,.
\end{equation}
The Coulomb interaction $\hat{\mathcal U}_{i}$ could
include crystal-field effects \cite{miyake2008,mozara2018}
but we assume a spherically symmetric operator
characterized by Slater parameters $F^i_k$, $k=0,2,\dots,2\ell_i$
\cite{CondonShortley}. The Coulomb interaction could act also
between shells, especially if they are located at the same atom
\cite{pivetta2020}, but we leave such settings out for now.


\paragraph*{Nickelocene-terminated STM tip above a Fe adatom.}
\addcontentsline{toc}{section}{Nickelocene-terminated STM tip above a Fe
  adatom}
We revisit the
inelastic tunneling spectra of an iron adatom on the Cu(100) surface
measured with an STM tip having a nickelocene (Nc) molecule attached to it
\cite{verlhac2019}. The nickel atom in nickelocene has 8 electrons
in its 3d shell carrying spin $S_\text{Ni}=1$. It has a
5-fold symmetry, thus only the axial anisotropy parameter $\DNc$ of
the anisotropic spin model can be nonzero, splitting the spin triplet into
a singlet, $S_{z,\text{Ni}}=0$ (the ground state), and a doublet,
$S_{z,\text{Ni}}=\pm1$. The magnitude of $\DNc$, which gives the gap
between the singlet and the doublet, slightly depends on the
environment around the Nc molecule, being between
$3.2$ and $4.0$~meV \cite{ormaza2017,ormaza2017a,verlhac2019,wackerlin2022}.
The motivation for attaching nickelocene to the STM tip is analogous
to magnetometry applications of the nitrogen-vacancy centers in
diamond \cite{degen2008,taylor2008}: a magnetic field or an exchange
field generated by nearby magnetic moments splits the
$S_{z,\text{Ni}}=\pm1$ doublet and when detected, this splitting
provides information about the field.

The Fe atom adsorbed at the hollow site of the Cu(100) surface carries
spin 
$S_\text{Fe}=3/2$ as found in density-functional-theory (DFT)
calculations \cite{verlhac2019,stepanyuk2003,suppl}. The site has
$C_{4v}$ symmetry which again allows only the axial anisotropy
$\DFe$ to be nonzero. The model of the coupled
Ni and Fe spins reads as
\begin{equation}
\label{eq:spin_model}
\hat{\mathcal H}_\text{spin} 
 = \DNc \hat S^2_{z,\text{Ni}} + \DFe \hat S^2_{z,\text{Fe}}
 - J \hat{\vec S}_\text{Ni}\cdot\hat{\vec S}_\text{Fe}
\end{equation}
when the isotropic Heisenberg exchange $J$ is assumed. In the setup of
Verlhac \emph{et al.} \cite{verlhac2019}, the Nc molecule is oriented
such that its $z$ axis is (nearly) perpendicular to
the Cu surface. The measured $d^2I/dV^2$ spectra show the Nc excitation
at 3.5~meV, the splitting of which increases as the STM tip approaches
the surface and the Ni--Fe distance decreases, and no other excitations are
detected up to 10 mV of applied voltage. The observations are well
reproduced by assuming the Fe atom to have a large out-of-plane
anisotropy, which effectively freezes the Fe spin to the
$S_{z,\text{Fe}}=\pm 3/2$ state in the whole voltage range
explored in the experiment. The two-spin model, Eq.~\eqref{eq:spin_model}, then
simplifies to a model for the Ni spin alone, with the exchange converted
to a Zeeman-like term $\Delta^{xc}_z \hat S_{z,\text{Ni}}$ with
$\Delta^{xc}_z=3J/2$. The $d^2I/dV^2$ spectra can be estimated
by methods reviewed in \cite{ternes2015} and their shape is controlled
by two
independent parameters: the exchange field $\Delta^{xc}_z$ gives the
splitting of the $S_{z,\text{Ni}}=\pm1$ level, and the polarization of
the surface electrode due to the frozen out-of-plane Fe spin
causes the asymmetry of the spectra with respect to
the voltage polarity \cite{loth2010}.

To apply our electronic model, we need to estimate
the parameters entering $\hat{\mathcal H}_{ns}$.
Ideally, they would be determined from first principles
\cite{ferron2015b,wolf2020} but we opt for a simpler heuristic
approach.
The spin-orbit parameter $\zeta$ and the Slater parameters
$F_k$ are taken from Hartree--Fock calculations \cite{cowanCode} of
Ni\textsuperscript{2+} ion
(3d\textsuperscript{8}4s\textsuperscript{0} configuration) and
Fe\textsuperscript{+} ion (3d\textsuperscript{7}4s\textsuperscript{0}
configuration). The parameters $F_k$ are additionally scaled
by a factor 0.8 to mimic screening
by the environment. The crystal-field parameters $B_{kq}$ are set such
that the individual orbitals in the ground state of
$\hat{\mathcal H}_{ns}$ are filled like in the DFT ground state and
the resulting spin anisotropy is consistent
with experiments. The Ni 3d shell in nickelocene has configuration
$(x^2-y^2,xy)^4(z^2)^2(xz,yz)^2$ \cite{swart2007}, and the
anisotropy is $\DNc=3.5$~meV. The 3d shell in Fe adatom
has configuration $(xy)^2(xz,yz)^3(z^2)^1(x^2-y^2)^1$ \cite{suppl},
and the anisotropy $\DFe$ is negative and
larger than 5~meV so that the first crystal-field excitation lies
above 10~meV (we use $\DFe=-7.5$~meV). This procedure does not specify
$B_{kq}$ uniquely, hence our calculations are not the
definite solution of the Fe adatom on Cu(100), rather, they represent
a proof of concept. The
hopping~$\hat{\mathcal{T}}_{ij}$, Eq.~\eqref{eq:T_operator}, is taken
in a simple form with $t_{mm'}~=~t \delta_{mm'}$
where $t$ is a monotonous decreasing function of the distance between
the Nc molecule and the Fe adatom. With such $\hat{\mathcal{T}}_{ij}$,
the spin~$\sigma$ and the magnetic quantum number $m$ are conserved
when electrons hop between the shells. All parameters
of~$\hat{\mathcal H}_{ns}$ are summarized in
Table~\ref{tbl:electron_parameters}.

\begin{figure}
\centering
\includegraphics[width=\linewidth]{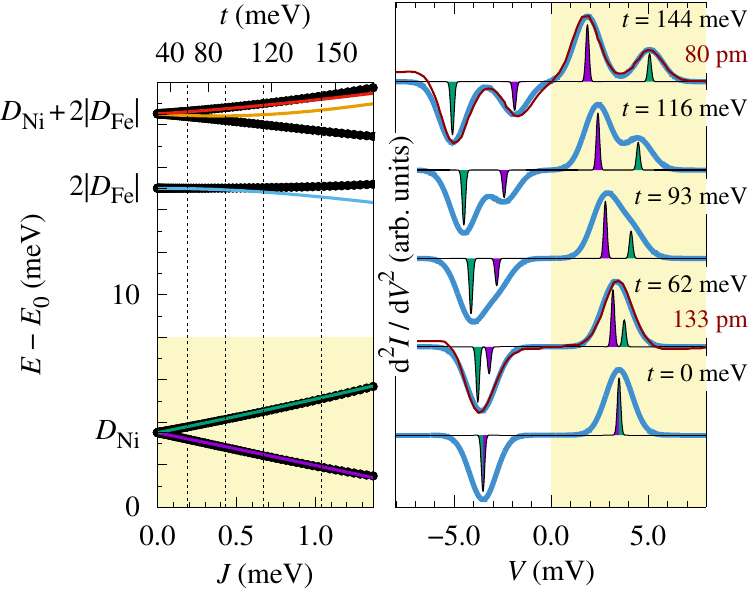}
\caption{\label{fig:spin_electron_model} Nickelocene-terminated tip
  atop a Fe adatom. Eigenstate energies (left~panel) from the
  electron model (thick black lines) and from the spin model (color
  lines) are shown as functions of the hopping~$t$ and the
  corresponding exchange~$J$.
  The $d^2I/dV^2$ spectra (right~panel) are calculated for values of
  $t$ indicated by vertical lines in the left panel.
  The colors of the peaks match the colors in the left panel.
  The thick blue lines are spectra calculated with an additional Gaussian
  broadening (1.6 meV FWHM). The thin red lines are the
  experimental spectra at tip--adatom distances $z=80$~pm and
  $z=133$~pm \cite{verlhac2019}, which are well reproduced with
  $t=144$~meV and $t=62$~meV.
  }
\end{figure}

The left panel of Fig.~\ref{fig:spin_electron_model} compares the
lowest excitations of $\hat{\mathcal H}_{ns}$ to the spectrum of the
spin model, Eq.~\eqref{eq:spin_model}, when the appropriate relation
between the hopping $t$ and the exchange $J$, $J\sim
t^2/U$, is utilized. The excitations around 3.5~meV, corresponding to
a change of the Ni spin direction, match perfectly. The
higher excitations, involving a change of the Fe spin direction,
deviate since the spin $3/2$ is not a
complete representation of the low-energy physics of the Fe atom in
the electron model -- there is an additional orbital degree of freedom
associated with the 3 electrons residing in the degenerate $(xz,yz)$ orbitals.
It could be an artifact of our approximation that
neglects the hybridization of the adatom with the
surface. Whether present or not, this degree of freedom does not
visibly influence the spectra below 10~mV (see End Matter).

The right panel of Fig.~\ref{fig:spin_electron_model} shows
$d^2I/dV^2$ spectra calculated from
Eqs.~\eqref{eqs:conductance} with $A_{i\Gamma}=A$
and $B_{i\Gamma}=0$ for $i\Gamma$ corresponding to the Ni atom,
and with $A_{i\Gamma}=0$ and $B_{i\Gamma}=B$ for $i\Gamma$
corresponding to the Fe atom. At large tip--adatom
distances (small hopping $t$), there is just a single peak at 3.5~mV
of applied voltage, which progressively splits as the distance
decreases and the interaction between the Fe and Ni atoms
increases. The individual spectra in
Fig.~\ref{fig:spin_electron_model} differ only in the value
of~$t$. The asymmetry with respect to the voltage polarity (the first
peak larger and the second smaller for positive voltages, and vice
versa for negative voltages) \emph{is a result of}
Eqs.~\eqref{eqs:conductance}. This is different
from the model applied in \cite{verlhac2019}, in which the
asymmetry was controlled by an additional parameter (the
polarization of the surface electrode). In the present case, when the
hopping $t$ is fitted to a particular splitting of the
nickelocene level, the intensities of the $d^2I/dV^2$ peaks
come out automatically and agree very well with the experimental data.


\paragraph*{Orbital magnetism.}
\addcontentsline{toc}{section}{Orbital magnetism}
In the investigations of adatoms with a significant orbital
contribution to their magnetic moments, the exchange interaction
with the tunneling electrons was traditionally assumed in the bilinear
Heisenberg form, either $\vec S\cdot\vec s$ \cite{schuh2011} or
$\vec J\cdot\vec s$ \cite{miyamachi2013,karlewski2015,balashov2018},
where $\vec{S}$ and $\vec{J}$ are the spin and total angular momentum of
the adatom, and $\vec s$ is the spin of the tunneling electron. A
consequence of this exchange is the selection rule $|\Delta
S_z|\leq 1$ or $|\Delta J_z|\leq 1$ for the transitions allowed in
the inelastic tunneling
\cite{hirjibehedin2007,fernandez-rossier2009}. It is known, however,
that exchange involving orbital moments is more 
complicated \cite{elliott1968}. Application of the Heisenberg exchange
to 4f
electrons in the context of the impurity models was criticized
already by Coqblin and Schrieffer \cite{coqblin1969}, and
deviations from the Heisenberg exchange were demonstrated also
experimentally, for instance in
Yb\textsubscript{2}Pt\textsubscript{2}Pb, where the $J_z=\pm7/2$ doublet 
behaves like $J_z=\pm1/2$, which would be impossible with a bilinear
exchange \cite{wu2016}.

\begin{figure}
\includegraphics[width=\linewidth]{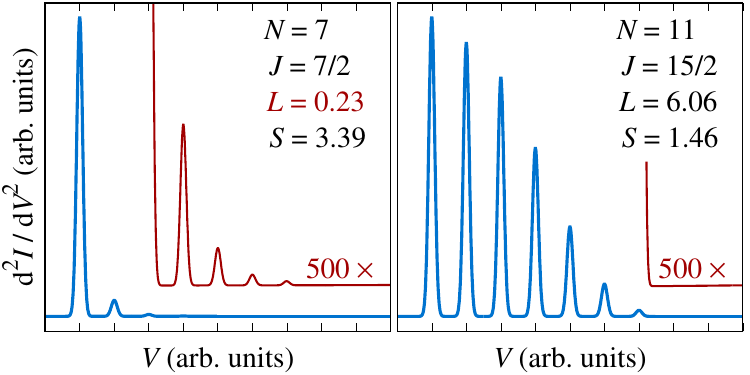}
\caption{\label{fig:fshellZeeman}$d^2I/dV^2$ spectrum of
an f~shell split by a magnetic field. The filling of the
shell is 7 (left) and 11 (right). The magnetism of the
shell with 7 electrons is dominated by spin, hence it displays only
a weak deviation from the ``standard'' selection rule $|\Delta
J_z|\leq 1$.}
\end{figure}

To visualize the selection rules predicted by our theory, we consider
$\hat{\cal H}_\text{ns}$ consisting of a single f shell with a strong
spin-orbit coupling corresponding to heavy lanthanides \cite{ogasawara1994}
or light actinides \cite{ogasawara1991}. The shell is split by a
magnetic field into $2J+1$ levels with
$J_z=J$ being the ground state. Figure~\ref{fig:fshellZeeman} shows
its $d^2I/dV^2$ spectra computed for $A_{i\Gamma}=A$ and
$B_{i\Gamma}=B$, which means the electrodes are spherically
symmetric around the shell (an artificial setup in the context of STM
but advantageous for the effect we intend to illustrate). As long as
$2J+1\ge 8$, there are seven transitions visible in the spectra, implying
the selection rule $|\Delta J_z|\leq 7$. This is consistent with the
arguments of \cite{coqblin1969} and follows from the maximal possible
change of the total angular momentum of the tunneling
electron, from $j_z=-7/2$ to $j_z=7/2$. (The electrode orbitals that
couple to the f shell have orbital momentum $\ell=3$
\cite{haldane1977,bringer1977,gunnarsson1989}, which together with the
spin adds up to $j=5/2$ and $j=7/2$.) For a general shell, the
selection rule is $|\Delta J_z|\leq 2\ell+1$.


\paragraph*{Summary.}
\addcontentsline{toc}{section}{Summary}
We developed a cotunneling model relevant for the inelastic electron
tunneling through nanosystems containing multiple magnetic
centers and/or for tunneling spectra measured with STM tips
functionalized with a magnetic center. We did not make any assumptions
about the magnetic exchange, in particular, it was not reduced to the
often-used bilinear form, and hence the model is applicable to
magnetic centers with large orbital moments (like lanthanide or
actinide atoms). The general form of the exchange and the consequent
less restrictive selection rules have implications also for the decay
rates of magnetic states of adatoms, since these rates are often
limited by scattering with the surface electrons.

\begin{acknowledgments}
This work was co-funded by the European Union and the Czech Ministry
of Education, Youth and Sports (Project TERAFIT --
{\text{CZ.02.01.01/00/22\_008/0004594}}). 
Computational resources were partly
provided by the e-INFRA CZ project (ID:90254), supported by the
Czech Ministry of Education, Youth and Sports.

The input files for the electronic structure calculations as well as
all data used to plot the figures in this Letter and in its
Supplemental Material \cite{suppl} are deposited at
\href{http://doi.org/10.5281/zenodo.16732408}{DOI:10.5281/zenodo.16732408}.
\end{acknowledgments}

\bibliography{IETS_HubI,suppl}


\newpage


\onecolumngrid
\section*{End Matter}
\twocolumngrid

\begin{figure}
\centering
\includegraphics[width=\linewidth]{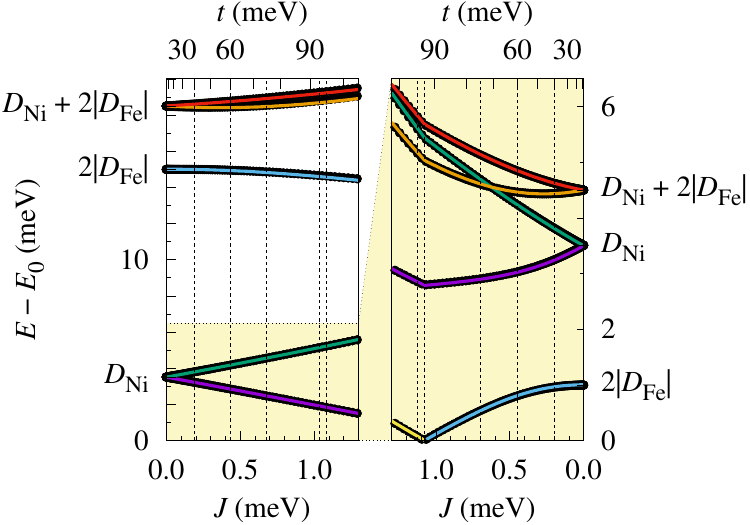}
\caption{Eigenstate energies calculated in the electron model (thick
black lines) and in the corresponding spin model (color lines) as
functions of the hopping $t$ and the corresponding exchange~$J$.  Two
scenarios of the anisotropy parameters are compared: $|\DFe|
> \DNc$ (left panel) and $|\DFe| < \DNc$ (right panel).  The vertical
lines indicate the $t$ values, for which the $d^2I/dV^2$ spectra are
shown in Fig.~\ref{fig:iets_spectra_diff_D}. Note the different ranges
of the vertical axes and the different directions of the horizontal
axes in the two panels.}
\label{fig:spin_electron_model_diff_D}
\end{figure} 

In the discussion of the IETS spectra of the Fe adatom on the Cu(100)
surface measured with a nickelocene-terminated STM tip, we assumed that
the Fe anisotropy is large so that the Fe spin is essentially frozen
in the $S_{z,\text{Fe}} = \pm 3/2$ state. This facilitated the
reduction of the two-spin model, Eq.~\eqref{eq:spin_model}, to a model
for the Ni spin only. Verlhac \emph{et al.} \cite{verlhac2019} allude
to this single-spin model being applicable also for small Fe
anisotropy as long as $|\DFe|>k_B T$. Here we test this conjecture by
explicitly comparing the predictions of our IETS theory in
two scenarios: $|\DFe| > \DNc$ and $|\DFe| < \DNc$.

We keep the parameters of the Ni 3d shell the same as before but we
alter the crystal field of the Fe 3d shell; we tune it
to give \hbox{$(xy)^2(x^2-y^2)^2(xz,yz)^2(z^2)^1$} as the
ground-state configuration, which
supports the out-of-plane easy axis like the DFT-based configuration
employed earlier. The numerical values of the Wybourne parameters $B_{kq}$,
corresponding to anisotropy parameters $\DFe=-7.5$~meV and
$\DFe=-0.5$~meV, are listed in Table~\ref{tbl:electron_parameters}.
 Figure~\ref{fig:spin_electron_model_diff_D} compares the
eigenstate energies computed in the electron model $\hat{\mathcal
  H}_{ns}$ and in the corresponding spin model $\hat{\mathcal H}_\text{spin}$,
Eq.~\eqref{eq:spin_model}. Unlike Fig.~\ref{fig:spin_electron_model},
they now match almost perfectly, since the present configuration of
the Fe 3d shell avoids the extra orbital degree of freedom associated
with the 3/4 filling of the doubly degenerate $(xz,yz)$
orbitals. Compared to Fig.~\ref{fig:spin_electron_model}, the
prefactor in the relation $J\sim t^2/U$ is now different (and it is
also slightly different in each of the two panels in
Fig.~\ref{fig:spin_electron_model_diff_D}).

\begin{figure}
\includegraphics[width=\linewidth]{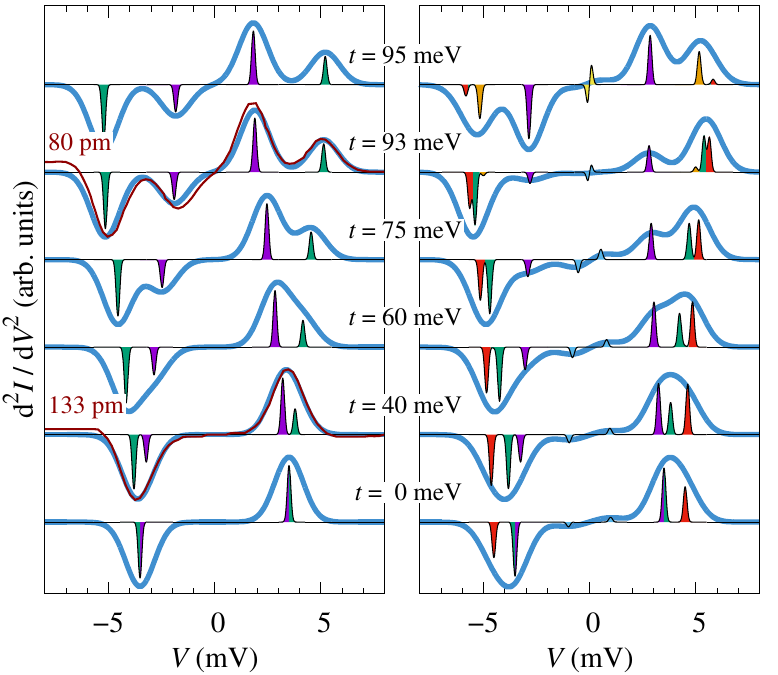}
\caption{Sets of $d^2I/dV^2$ spectra calculated for a
nickelocene-terminated tip atop the Fe adatom assuming $|\DFe| > \DNc$
(left panel) and $|\DFe| < \DNc$ (right panel). The colors
of the peaks match the colors in
Fig.~\ref{fig:spin_electron_model_diff_D}. The orange peak in the
right panel has a vanishingly small intensity for $t\leq 75$~meV but
it is not forbidden. The
thick blue lines are spectra calculated with an additional Gaussian broadening
(1.6 meV FWHM). The thin red lines in the left panel are the
experimental data obtained at tip--adatom distances $z=80$~pm and
$z=133$~pm \cite{verlhac2019}. The two topmost spectra in the right panel
are taken just before and just after the ground state and the first
excited state interchange.}
\label{fig:iets_spectra_diff_D}
\end{figure}

Figure~\ref{fig:iets_spectra_diff_D} shows
$d^2I/dV^2$ spectra calculated from
Eqs.~\eqref{eqs:conductance} in the same way as
spectra in Fig.~\ref{fig:spin_electron_model}. The present spectra
obtained for $|\DFe| > \DNc$ are almost indistinguishable
from the spectra in Fig.~\ref{fig:spin_electron_model} despite being
calculated with a substantially different crystal-field potential. It
is not so surprising, however, since the Fe degrees of freedom are
frozen in the same spin state $S_{z,\text{Fe}} = \pm 3/2$ in the whole
voltage range plotted in the figures, and these spectra are thus only
a very limited probe of the Fe crystal field.

\begin{figure}
\includegraphics[width=\linewidth]{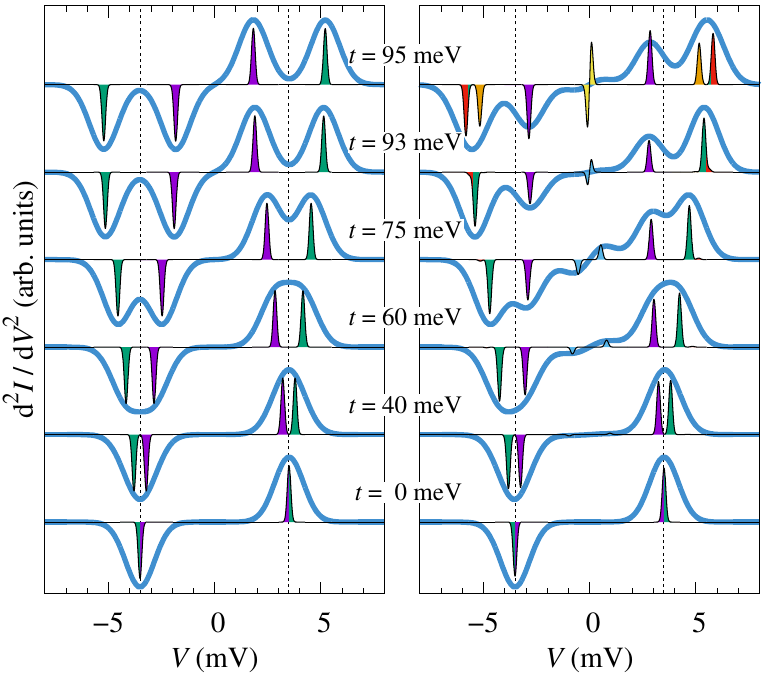}
\caption{Sets of $d^2I/dV^2$ spectra calculated for a
nickelocene-terminated tip atop the Fe adatom assuming $|\DFe| > \DNc$
(left panel) and $|\DFe| < \DNc$ (right panel), when the tunneling
current does not pass through the Fe 3d shell. Everything else is the
same as in Fig.~\ref{fig:iets_spectra_diff_D}.}
\label{fig:iets_spectra_diff_D_onlyNi}
\end{figure}

\def\m{\phantom{-}}
\def\0{\phantom{0}}

\begin{table*}
\caption{\label{tbl:electron_parameters}
Parameters of the atomic shells constituting our illustrative nanosystems
$\hat{\mathcal H}_{ns}$, and the axial anisotropy parameter~$D$ of the
spin models corresponding to these shells. Rows A and B are used
in Fig.~\ref{fig:spin_electron_model}, rows A, C and~D in
Figs.~\ref{fig:spin_electron_model_diff_D}--\ref{fig:iets_spectra_diff_D_onlyNi},
and rows E and~F in Fig.~\ref{fig:fshellZeeman}. The
origin of the parameters of the 3d shells is discussed in the text,
the parameters of the f~shells are taken from \cite{ogasawara1991}
(U\textsuperscript{3+} ion), the magnetic field $B_z$ is chosen such
that the Zeeman level spacing comes out as 1~meV.
}
\renewcommand{\arraystretch}{1.3}
\centering
\begin{ruledtabular}
\begin{tabular}{llcccccccccccc}
 & ion & configuration
 & $D$ & $\zeta$ & $\epsilon=B_{00}$ & $B_{20}$ & $B_{40}$
 & $B_{44}$ & $B_z$ & $F_0$ & $F_2$ & $F_4$ & $F_6$\\
 &&& (meV) & (meV) & (eV) & (eV) & (eV) & (eV) & (T) & (eV) & (eV) & (eV) & (eV)\\
\hline
A & Ni\textsuperscript{2+}
  & $(x^2-y^2,xy)^4(z^2)^2(xz,yz)^2$
  & $\m3.5$ 
  & 82.6 & $-58.6$ & 3.400 & $-8.160$  & & & 8.0 & 9.786 & 6.078\\
B & Fe\textsuperscript{+}
  & $(xy)^2(xz,yz)^3(z^2)^1(x^2-y^2)^1$
  & $-7.5$ 
  & 45.4 & $-50.6$ & 0.140 & $\m0.646$ & $-1.762$ & & 8.0 & 7.809 & 4.814\\
\hline
C & Fe\textsuperscript{+}
  & $(xy)^2(x^2-y^2)^2(xz,yz)^2(z^2)^1$
  & $-7.5$ 
  & 45.4 & $-50.6$ & 7.840 & $\m2.184$ & $\m0.000$ & & 8.0 & 7.809 & 4.814\\
D & Fe\textsuperscript{+}
  & $(xy)^2(x^2-y^2)^2(xz,yz)^2(z^2)^1$
  & $-0.5$ 
  & 45.4 & $-50.6$ & 4.700 & $-1.200$  & $\m0.000$ & & 8.0 & 7.809 & 4.814\\
\hline
E & $n$f\textsuperscript{7}  &&
  & 235  & $-17.95$ & & & & $\08.80$ & 3.0 & 7.09\0 & 4.60\0 & 3.36\0 \\
F & $n$f\textsuperscript{11} &&
  & 235  & $-29.90$ & & & &  $14.48$ & 3.0 & 7.09\0 & 4.60\0 & 3.36\0 \\
\end{tabular}
\end{ruledtabular}
\end{table*}

The $d^2I/dV^2$ spectra corresponding to the case $|\DFe| < \DNc$ are
very different since the excitations involving the Ni and Fe spins are
mixed together. Their shape is clearly incompatible with the
experimental data \cite{verlhac2019}. These spectra are, however,
similar to the spectra measured with a nickelocene-terminated tip on
Cr atoms in metal-organic polymers having a spin $S_\text{Cr}=2$,
out-of-plane easy axis, and a small anisotropy parameter
$D_\text{Cr}\approx -0.23$~meV \cite{wackerlin2022}.

To bring the spectra of the scenario with $|\DFe| < \DNc$ closer
to the experiment, there would have to be a mechanism suppressing the
excitations involving the Fe spin. Essentially, the tunneling
electrons would have to travel along a path that avoids the Fe 3d
shell, passing instead through some orbital(s) also centered at the Fe
atom but being more diffuse than the 3d shell, for instance the 4s
orbitals. Being more diffuse implies that these orbitals should be
substantially hybridized with the electronic states of the surface.
The simplest description we can build amounts to considering these
extra orbitals as a part of the surface electrode, completely
decoupling the Fe 3d shell from the surface, and introducing tunneling
from the surface directly to the Ni 3d shell. This means setting
$A_{i\Gamma}=A$ and $B_{i\Gamma}=B$ for $i\Gamma$ corresponding to the
Ni atom, and $A_{i\Gamma}=0$ and $B_{i\Gamma}=0$ for $i\Gamma$
corresponding to the Fe atom.

The $d^2I/dV^2$ spectra calculated in such an alternative tunneling setup
are shown in Fig.~\ref{fig:iets_spectra_diff_D_onlyNi}. The Fe-related
excitations are indeed suppressed and the spectra reflect mostly just the
nickelocene level split by the exchange field in both scenarios,
$|\DFe| > \DNc$ and $|\DFe| < \DNc$. The peak intensities, however,
are not consistent with the experiment -- the spectra are
antisymmetric with respect to the polarity of the tip-surface
voltage~$V$. Since the tunneling current now does not pass through the
Fe 3d shell, no equivalent of the apparent polarization of the surface
electrode is generated and hence the mechanism causing the spectra
asymmetry is lost. Another aspect making the $|\DFe| < \DNc$ case
inconsistent with the experimental data is the asymmetric
splitting of the nickelocene level around its energy at large
tip--adatom distances ($t=0$~meV) apparent already in
Fig.~\ref{fig:spin_electron_model_diff_D}.

Putting all these observations together, we can conclude that the IETS
spectra reported by Verlhac \emph{et al.} \cite{verlhac2019} represent an
evidence for the anisotropy $\DFe$ of the Fe adatom at the hollow site
of the Cu(100) surface to be negative and larger than
5~meV. Precise determination of the actual magnitude of $\DFe$ would
require an experimental detection of the excitations corresponding to
changes of the Fe spin direction.


\end{document}



\begin{center}
{\bfseries Supplementary Information for\par}
\vskip.8ex

{\large\bfseries Inelastic Electron Tunneling through Adatoms and
  Molecular Nanomagnets\par}

\vskip 1em
Daria K\'yvala and Jind\v{r}ich Koloren\v{c}\footnote{kolorenc@fzu.cz}\\
{\itshape Institute of Physics (FZU), Czech Academy of Sciences, Na
  Slovance 2, 182~00~Praha 8, Czech Republic}
\end{center}

\hrule
\tableofcontents
\vskip\baselineskip
\hrule


\section{Summary of notation}

Numbers of sections, figures, tables, equations and references in this
supplemental material are prefixed with letter ``S''. Numbers without prefix
refer to the figures and equations from the main article.

{
\setlength{\columnsep}{2.3em}
\begin{multicols}{2}
\raggedright
\begin{itemize}\itemsep0pt
\item[$\epsilon$] one-particle energies
\item[$\epsilon_\text{F}$] Fermi energy of the surface
\item[$E$] many-body energies of the nanosystem, tip and surface
\item[$\mathcal E$] energies of the whole system
\item[$i$] index of the magnetic center in the nanosystem
\item[$m$, $n$, $M$] magnetic quantum number
\item[$\sigma$] spin quantum number
\item[$\Gamma$] label of the irreducible representation at a magnetic
center
\item[$\mu$] label of a component of $\Gamma$ if it is multidimensional;
$\mu\equiv m\sigma$ for a spherically symmetric site without spin-orbit
coupling
\item[$k$] a star of a wave vector combined with the band index for
the conduction states in tip and surface
\item[$\hat d^\dagger$, $\hat d$] creation and annihilation operators
at magnetic centers in the nanosystem
\item[$\hat a^\dagger$, $\hat a$] creation and annihilation operators
in the tip
\item[$\hat b^\dagger$, $\hat b$] creation and annihilation operators
in the surface
\item[$N$] ground-state number of electrons in the nanosystem
\item[$|\psi\rangle$] many-body eigenstates of the nanosystem
\item[$\alpha$, $\beta$, $\gamma$] labels of the eigenstates $|\psi\rangle$ for
a given filling $N$ of the nanosystem
\item[$V$] tip--surface bias voltage
\item[$|\phi_s\rangle$] ground state of the surface, one-particle
levels filled up to~$\epsilon_\text{F}$
\item[$|\phi_t\rangle$] ground state of the tip, one-particle
levels filled up to the energy $\epsilon_\text{F}+eV$
\item[$E_s$] energy of the state $|\phi_s\rangle$
\item[$E_t$] energy of the state $|\phi_t\rangle$
\item[$|\chi\rangle$] states of the whole system in the form of
a direct product $|\phi_t\rangle \otimes |\phi_s\rangle \otimes
|\psi\rangle$
\item[$|v\rangle$] intermediate virtual states of the tunneling process
\item[$|f\rangle$] final states of the tunneling process
\item[$\hat{\mathcal H}$] Hamiltonian and its parts
\item[$\hat{\mathcal V}$] tunneling operator
(tip/surface${}\leftrightarrow{}$nanosystem)
\item[$\hat{\mathcal T}$] hopping operator connecting individual
magnetic centers in the nanosystem
\item[$\hat{\mathcal O}$] operator giving amplitudes of inelastic
transitions in the nanosystem induced by an electron tunneling through it
\item[$A$] tip${}\leftrightarrow{}$nanosystem tunneling amplitude
\item[$B$] surface${}\leftrightarrow{}$nanosystem tunneling amplitude
\item[$t$] amplitude of hopping between individual centers
\item[$S$, $L$, $J$] spin, orbital
and total moment of a magnetic center
\item[$g$] differential conductance
\item[$\rho$] density of states in the tip and surface
\end{itemize}
\end{multicols}
}




\section{Derivation of the differential conductance}
\label{sec:dIdV_derivation}

\begin{figure}
\centering
\includegraphics[width=0.82\linewidth]{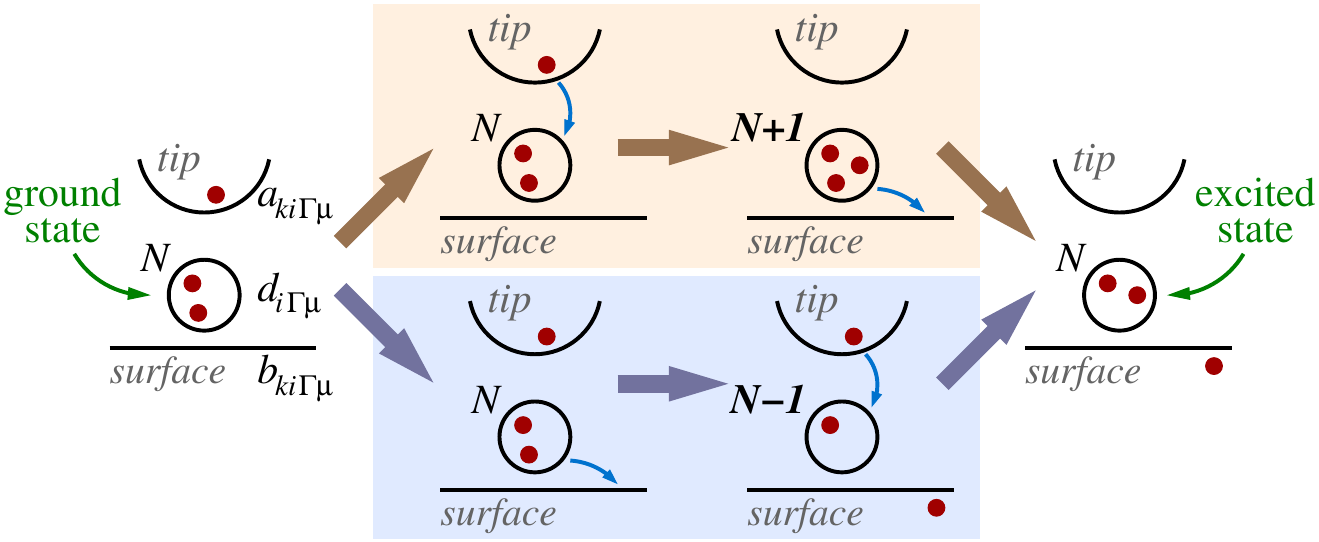}
\caption{\label{fig:cotunneling}Two contributions to
  the positive current $I_+$: an electron tunnels from the tip to
  the surface (top path) and a hole tunnels from the surface to the tip
  (bottom path).}
\label{fig:tunnel_one_by_one}
\end{figure}

We are concerned with arrangements composed of two electrodes (STM tip and
the surface) and a magnetic nanosystem placed in between the
electrodes. The electrodes are assumed to host non-interacting
electrons, the nanosystem is a general many-body system.

The contribution to the tunneling current from processes that start
when the magnetic nanosystem resides in an $N$-electron state
$|\psi_{N\alpha}\rangle$ is given by the Kramers--Heisenberg formula
\begin{equation*}
\tag{\ref*{eq:KHformula}} 
\label{eq:KHformulaSuppl}
I_{\pm}^{(\alpha)} = \pm \frac{2\pi e}{\hbar}\sum_{f_{\pm}}\biggl|
\sum_{v}\frac{\langle f_\pm|\hat{\mathcal V}|v\rangle
\langle v|\hat{\mathcal V}|\chi_{N\alpha}\rangle}{%
\mathcal E_{v}-\mathcal E_{N\alpha}}\biggr|^2
\delta\bigl(\mathcal E_{f_\pm}-\mathcal E_{N\alpha}\bigr)\,,
\end{equation*}
where the positive current $I_+$ comes from the final states
\begin{subequations}
\label{eq:finalStates}
\begin{equation}
|f_+\rangle = \hat b^{\dagger}_{k'i'\Gamma'\mu'}
\hat a_{ki\Gamma\mu}|\chi_{N\beta}\rangle
\quad\text{with}\quad \ek < \eF+eV\text{ and } \ekk > \eF\,,
\end{equation}
and the negative current $I_-$ from the final states
\begin{equation}
|f_-\rangle = \hat a^{\dagger}_{ki\Gamma\mu}
\hat b_{k'i'\Gamma'\mu'} |\chi_{N\beta}\rangle
\quad\text{with}\quad \ek > \eF+eV\text{ and } \ekk < \eF\,.
\end{equation}
\end{subequations}
The final states are written with the help of product states
$|\chi_{N\beta}\rangle=|\phi_t\rangle \otimes |\phi_s\rangle \otimes
|\psi_{N\beta}\rangle$, where $|\phi_t\rangle$ and $|\phi_s\rangle$
are ground states of the tip and surface electrodes (each of which is
a Fermi see filled up to the corresponding Fermi level). The
final-state energies then are
\begin{subequations}
\begin{align}
\mathcal E_{f_+}&=E_{N\beta}+\Et+\Es-\ek+\ekk\,,\\
\mathcal E_{f_-}&=E_{N\beta}+\Et+\Es+\ek-\ekk\,,
\end{align}
\end{subequations}
where $E_{N\beta}$, $\Et$ and $\Es$ are energies of
$|\psi_{N\beta}\rangle$, $|\phi_t\rangle$ and $|\phi_s\rangle$.
As discussed in the main text, the tunneling Hamiltonian
$\hat{\mathcal{V}}$ has the form
\begin{equation*}
\hat{\mathcal{V}} = \sum_{ki\Gamma\mu} \Bigl(
A_{ki\Gamma} \hat{a}^\dagger_{ki\Gamma\mu}\hat{d}_{i\Gamma\mu}
+ A_{ki\Gamma}^* \hat{d}_{i\Gamma\mu}^\dagger \hat{a}_{ki\Gamma\mu}
+ B_{ki\Gamma} \hat{b}^{\dagger}_{ki\Gamma\mu}  \hat{d}_{i\Gamma\mu}
+ B_{ki\Gamma}^* \hat{d}_{i\Gamma\mu}^\dagger \hat{b}_{ki\Gamma\mu}
\Bigr),
\tag{\ref*{eq:coupling}} 
\label{eq:couplingSuppl}
\end{equation*}
which implies that each final state can be reached by two paths, that
is, via two intermediate states: one with $N+1$ electrons and the
other with $N-1$ electrons in the nanosystem (see
Fig.~\ref{fig:tunnel_one_by_one}). The combinations of the
intermediate states $|v\rangle$ and the final states $|f_\pm\rangle$
that contribute to the Kramers--Heisenberg formula,
Eq.~\eqref{eq:KHformulaSuppl}, are summarized in
Table~\ref{tab:KHcontrib}. We will work out the partial expressions
corresponding to the individual rows of the table separately.

\begin{table}
\caption{\label{tab:KHcontrib}Combinations of intermediate and
  final states that give non-vanishing contributions in the
  Kramers--Heisenberg formula, Eq.~\eqref{eq:KHformulaSuppl}. The
  restrictions on $k$ and $k'$ that apply to the intermediate states
  are the same as the restrictions that apply to the corresponding final
  states and that are listed in Eqs.~\eqref{eq:finalStates}.}
\vskip.4em
\def\arraystretch{1.4}
\begin{tabular*}{\textwidth}{c@{\extracolsep{\fill}}cc}
\hline\hline
final state & intermediate state $|v\rangle$ &
intermediate energy $\mathcal E_v$ \\
\hline
\multirow{2}{*}{$|f_+\rangle = \hat b^{\dagger}_{k'i'\Gamma'\mu'}
\hat a_{ki\Gamma\mu}|\chi_{N\beta}\rangle$} &
$\hat a_{ki\Gamma\mu}|\chi_{N+1,\gamma}\rangle$ &
$E_{N+1,\gamma} + \Et + \Es - \ek$ \\
%
&
$\hat b^{\dagger}_{k'i'\gamma'\mu}|\chi_{N-1,\gamma}\rangle$ &
$E_{N-1,\gamma} + \Et + \Es + \ekk$  \\
\hline
\multirow{2}{*}{$|f_-\rangle = \hat a^{\dagger}_{ki\Gamma\mu}
\hat b_{k'i'\Gamma'\mu'}|\chi_{N\beta}\rangle$} &
$\hat b_{k'i'\Gamma'\mu'\sigma'}|\chi_{N+1,\gamma}\rangle$ &
$E_{N+1,\gamma} + \Et + \Es - \ekk$  \\
%
&
$\hat a^{\dagger}_{ki\Gamma\mu}|\chi_{N-1,\gamma}\rangle$ &
$E_{N-1,\gamma} + \Et + \Es + \ek$ \\
\hline\hline
\end{tabular*}%
\end{table}



\subsection{Positive current}

The first half of the sum over the intermediate states $|v\rangle$,
which involves the intermediate states with an extra electron in the
nanosystem, can be consecutively rewritten as
\begin{multline}
\label{eq:sum1IntermStatesPos}
\sideset{}{^{(1)}}\sum_v
\frac{\langle f_+|\hat{\mathcal V}|v\rangle
\langle v|\hat{\mathcal V}|\chi_{N\alpha}\rangle}{%
\mathcal E_{v}-\mathcal E_{N\alpha}}=\\
\sum_\gamma \frac{\langle \chi_{N\beta}|
\hat a^\dagger_{ki\Gamma\mu}\hat b_{k'i'\Gamma'\mu'}\hat{\mathcal V}
\hat a_{ki\Gamma\mu}|\chi_{N+1,\gamma}\rangle
\langle \chi_{N+1,\gamma}|\hat a_{ki\Gamma\mu}^\dagger
\hat{\mathcal V}|\chi_{N\alpha}\rangle}{%
E_{N+1,\gamma}-E_{N\alpha} - \ek}=\\
%
%
B_{k'i'\Gamma'}A^*_{ki\Gamma}\sum_\gamma \frac{%
\langle \psi_{N\beta}|\hat d_{i'\Gamma'\mu'}|\psi_{N+1,\gamma}\rangle
\langle \psi_{N+1,\gamma}|\hat d^{\dagger}_{i\Gamma\mu}
|\psi_{N\alpha}\rangle}{%
E_{N+1,\gamma}-E_{N\alpha} - \ek}=\\
B_{k'i'\Gamma'}A^*_{ki\Gamma}
\langle \psi_{N\beta}|\hat d_{i'\Gamma'\mu'}
\frac1{\hat{\mathcal H}_{ns}-E_{N\alpha} - \ek}
\hat d^{\dagger}_{i\Gamma\mu}
|\psi_{N\alpha}\rangle\,.
\end{multline}
In the first step, we substituted the appropriate expressions for
$|f_+\rangle$, $|v\rangle$ and $\mathcal E_v$. In the second step, we
inserted the tunneling operator $\hat{\mathcal V}$ and evaluated the
matrix elements in the subspace of tip and substrate under the
assumption that all single-particle states $\hat a_{ki\Gamma\mu}$ and
$\hat b_{ki\Gamma\mu}$ are mutually orthogonal. The matrix elements
are then either $+1$ or $-1$ depending on the order of the fermionic
operators. Finally, we realized that the sum over $\gamma$ is a sum
over the complete basis of the nanosystem states with $N+1$ electrons.

The second half of the sum over the intermediate states $|v\rangle$,
which involves the intermediate states with an extra hole in the
adatom, can be analogously transformed to a similar form,
\begin{multline}
\sideset{}{^{(2)}}\sum_v
\frac{\langle f_+|\hat{\mathcal V}|v\rangle
\langle v|\hat{\mathcal V}|\chi_{N\alpha}\rangle}{%
\mathcal E_{v}-\mathcal E_{N\alpha}}=\\
\sum_\gamma \frac{\langle \chi_{N\beta}|
\hat a^\dagger_{ki\Gamma\mu}\hat b_{k'i'\Gamma'\mu'}\hat{\mathcal V}
\hat b^{\dagger}_{k'i'\Gamma'\mu'}|\chi_{N-1,\gamma}\rangle
\langle\chi_{N-1,\gamma}|\hat b_{k'i'\Gamma'\mu'}
\hat{\mathcal V}|\chi_{N\alpha}\rangle}{%
E_{N-1,\gamma}-E_{N\alpha} + \ekk}=\\
%
%
-A^*_{ki\Gamma}B_{k'i'\Gamma'}\sum_\gamma \frac{%
\langle \psi_{N\beta}|\hat d^{\dagger}_{i\Gamma\mu}
|\psi_{N-1,\gamma}\rangle
\langle\psi_{N-1,\gamma}|\hat d_{i'\Gamma'\mu'}
|\psi_{N\alpha}\rangle}{%
E_{N-1,\gamma}-E_{N\alpha} + \ekk}=\\
-A^*_{ki\Gamma}B_{k'i'\Gamma'}
\langle \psi_{N\beta}|\hat d^{\dagger}_{i\Gamma\mu}
\frac1{\hat{\mathcal H}_{ns}-E_{N\alpha} + \ekk}
\hat d_{i'\Gamma'\mu'}|\psi_{N\alpha}\rangle\,.
\end{multline}
Substituting the above two partial expressions to Eq.~\eqref{eq:KHformulaSuppl}
yields a formula for the positive tunneling current,
\begin{multline}
\label{eq:posCurrentKSum}
I_+^{(\alpha)} = \frac{2\pi e}{\hbar}\hskip-2em\sum_{\substack{%
\beta kk'ii'\Gamma\Gamma'\mu\mu'\\[.2em]
\ek < \eF+eV\text{ and } \ekk > \eF}}\hskip-2em
|A_{ki\Gamma}B_{k'i'\Gamma'}|^2
\biggl|
\langle \psi_{N\beta}|\hat d_{i'\Gamma'\mu'}
\frac1{\hat{\mathcal H}_{ns}-E_{N\alpha} - \ek}
\hat d^{\dagger}_{i\Gamma\mu}
|\psi_{N\alpha}\rangle\\
-
\langle \psi_{N\beta}|\hat d^{\dagger}_{i\Gamma\mu}
\frac1{\hat{\mathcal H}_{ns}-E_{N\alpha} + \ekk}
\hat d_{i'\Gamma'\mu'}|\psi_{N\alpha}\rangle
\biggr|^2
\delta(\Delta_{\beta\alpha}-\ek+\ekk)\,,
\end{multline}
where the sum runs over all parameters of the final state
$|f_+\rangle$ and where a difference (the excitation energy)
$\Delta_{\beta\alpha}=E_{N\beta}-E_{N\alpha}$ was introduced. If the
$k$~dependence of the tunneling amplitudes $A$ and $B$ can be
neglected, the only remaining dependence on $k$ and $k'$ is through
the corresponding single-particle energies and hence these sums can be
rewritten as integrals weighted with the density of states in the tip,
$\rho^t(\epsilon)$, and in the surface, $\rho^s(\epsilon')$
\begin{multline}
I_+^{(\alpha)} = \frac{2\pi e}{\hbar} \sum_{\beta ii'\Gamma\Gamma'\mu\mu'}
|A_{i\Gamma}B_{i'\Gamma'}|^2
\int_{-\infty}^{eV}\rmd\epsilon\, \rho_{i\Gamma}^t(\epsilon)
\int_{0}^{\infty}\rmd\epsilon'\,\rho_{i'\Gamma'}^s(\epsilon')\,
\delta(\Delta_{\beta\alpha}-\epsilon+\epsilon')\\
\times
\biggl|
\langle \psi_{N\beta}|\hat d_{i'\Gamma'\mu'}
\frac1{\hat{\mathcal H}_{ns} - E_{N\alpha} - \epsilon}
\hat d^{\dagger}_{i\Gamma\mu}
|\psi_{N\alpha}\rangle\\
-
\langle \psi_{N\beta}|\hat d^{\dagger}_{i\Gamma\mu}
\frac1{\hat{\mathcal H}_{ns} - E_{N\alpha} + \epsilon'}
\hat d_{i'\Gamma'\mu'}|\psi_{N\alpha}\rangle
\biggr|^2\,.
\end{multline}
In this formula, we make $\eF$ the reference energy and set $\eF=0$.
Since the voltage~$V$ enters only in the upper limit of the integral over
$\epsilon$, it is now straightforward to evaluate the differential
conductance, $g_+^{(\alpha)}=\rmd I_+^{(\alpha)}/\rmd V$, which yields
\begin{multline}
g_+^{(\alpha)} = \frac{2\pi e^2}{\hbar} \sum_{\beta ii'\Gamma\Gamma'\mu\mu'}
|A_{i\Gamma}B_{i'\Gamma'}|^2\rho_{i\Gamma}^t(eV)
\int_{0}^{\infty}\rmd\epsilon'\,\rho_{i'\Gamma'}^s(\epsilon')\,
\delta(\Delta_{\beta\alpha}-eV+\epsilon')\\
\times
\biggl|
\langle \psi_{N\beta}|\hat d_{i'\Gamma'\mu'}
\frac1{\hat{\mathcal H}_{ns} - E_{N\alpha} - eV}
\hat d^{\dagger}_{i\Gamma\mu}
|\psi_{N\alpha}\rangle\\
-
\langle \psi_{N\beta}|\hat d^{\dagger}_{i\Gamma\mu}
\frac1{\hat{\mathcal H}_{ns} - E_{N\alpha} + \epsilon'}
\hat d_{i'\Gamma'\mu'}|\psi_{N\alpha}\rangle
\biggr|^2\,.
\end{multline}
Finally, the integral over $\epsilon'$ contributes only if
$\Delta_{\beta\alpha}-eV<0$ and hence we arrive at the expression
\begin{multline}
g_+^{(\alpha)} = \frac{2\pi e^2}{\hbar} \hskipBSum
\sum_{\substack{\beta\\[.2em]\Delta_{\beta\alpha}<eV}}\hskipBSum
\sum_{ii'\Gamma\Gamma'\mu\mu'}
|A_{i\Gamma}B_{i'\Gamma'}|^2
\rho_{i\Gamma}^t(eV)\rho_{i'\Gamma'}^s(eV-\Delta_{\beta\alpha})
\\
\times
\biggl|
\langle \psi_{N\beta}|\hat d_{i'\Gamma'\mu'}
\frac1{\hat{\mathcal H}_{ns} - E_{N\alpha} - eV}
\hat d^{\dagger}_{i\Gamma\mu}
|\psi_{N\alpha}\rangle\\
-
\langle \psi_{N\beta}|\hat d^{\dagger}_{i\Gamma\mu}
\frac1{\hat{\mathcal H}_{ns} - E_{N\beta} + eV}
\hat d_{i'\Gamma'\mu'}|\psi_{N\alpha}\rangle
\biggr|^2\,.
\end{multline}
To allow for a more compact presentation, we introduce a transition operator
\begin{equation*}
\tag{\ref*{eq:transOp}} 
\label{eq:transOpSuppl}
\hat{\mathcal O}^{\alpha\beta}_{i\Gamma\mu\,\, i'\Gamma'\mu'}
=
\colorbox{orange_bg}{$\displaystyle
\hat d_{i'\Gamma'\mu'}
\frac1{\hat{\mathcal H}_{ns} - E_{N\alpha} - eV}
\hat d^{\dagger}_{i\Gamma\mu}%
$}
-
\colorbox{blue_bg}{$\displaystyle
\hat d^{\dagger}_{i\Gamma\mu}
\frac1{\hat{\mathcal H}_{ns} - E_{N\beta} + eV}
\hat d_{i'\Gamma'\mu'}%
$}
\,,
\end{equation*}
with the help of which the differential conductance becomes
\begin{multline}
\label{eq:gPlusFinal}
g_+^{(\alpha)} = \frac{2\pi e^2}{\hbar}\hskipBSum
\sum_{\substack{\beta\\[.2em]\Delta_{\beta\alpha}<eV}}\hskipBSum
\sum_{ii'\Gamma\Gamma'\mu\mu'}
|A_{i\Gamma}B_{i'\Gamma'}|^2
\rho_{i\Gamma}^t(eV)\rho_{i'\Gamma'}^s(eV-\Delta_{\beta\alpha})
\\[-1em]
\times
\bigl|\langle \psi_{N\beta}|
\hat{\mathcal O}^{\alpha\beta}_{i\Gamma\mu\,\, i'\Gamma'\mu'}
|\psi_{N\alpha}\rangle\bigr|^2\,.
\end{multline}
The color boxes in Eq.~\eqref{eq:transOpSuppl} mark the two
cotunneling channels shown in Fig.~\ref{fig:cotunneling}.




\subsection{Negative current}

The derivation of a formula for the negative current and the
corresponding differential conductance follows the same steps as the
derivation of the positive current. The two parts
of the sum over the intermediate states in Eq.~\eqref{eq:KHformulaSuppl}
read as
\begin{multline}
\sideset{}{^{(1)}}\sum_v
\frac{\langle f_-|\hat{\mathcal V}|v\rangle
\langle v|\hat{\mathcal V}|\chi_{N\alpha}\rangle}{%
\mathcal E_{v}-\mathcal E_{N\alpha}}=\\
\sum_\gamma \frac{\langle \chi_{N\beta}|
\hat b_{k'i'\Gamma'\mu'}^\dagger \hat a_{ki\Gamma\mu}\hat{\mathcal V}
\hat b_{k'i'\Gamma'\mu'}|\chi_{N+1,\gamma}\rangle
\langle\chi_{N+1,\gamma}|\hat b_{k'i'\Gamma'\mu'}^\dagger
\hat{\mathcal V}|\chi_{N\alpha}\rangle}{%
E_{N+1,\gamma}-E_{N\alpha}- \ek}=\\
%
%
A_{ki\Gamma}B^*_{k'i'\Gamma'}\sum_\gamma \frac{%
\langle \psi_{N\beta}|\hat d_{i\Gamma\mu}
|\psi_{N+1,\gamma}\rangle
\langle\psi_{N+1,\gamma}|\hat d^{\dagger}_{i'\Gamma'\mu'}
|\psi_{N\alpha}\rangle}{%
E_{N+1,\gamma}-E_{N\alpha} - \ekk}=\\
A_{ki\Gamma}B^*_{k'i'\Gamma'}
\langle \psi_{N\beta}|\hat d_{i\Gamma\mu}
\frac1{\hat{\mathcal H}_{ns}-E_{N\alpha} - \ekk}
\hat d^{\dagger}_{i'\Gamma'\mu'}
|\psi_{N\alpha}\rangle
\end{multline}
and
\begin{multline}
\sideset{}{^{(2)}}\sum_v
\frac{\langle f_-|\hat{\mathcal V}|i\rangle
\langle i|\hat{\mathcal V}|\chi_{N\alpha}\rangle}{%
\mathcal E_{v}-\mathcal E_{N\alpha}}=\\
\sum_\gamma \frac{\langle \chi_{N\beta}|
\hat s_{k'i'\Gamma'\mu'}^\dagger \hat t_{ki\Gamma\mu}\hat{\mathcal V}
\hat t^{\dagger}_{ki\Gamma\mu}|\chi_{N-1,\gamma}\rangle
\langle\chi_{N-1,\gamma}|\hat t_{ki\Gamma\mu}
\hat{\mathcal V}|\chi_{N\alpha}\rangle}{%
E_{N-1,\gamma}-E_{N\alpha} + \ek}=\\
%
%
-B^*_{k'i'\Gamma'}A_{ki\Gamma}\sum_\gamma \frac{\langle \psi_{N\beta}|
\hat d^{\dagger}_{i'\Gamma'\mu'}|\psi_{N-1,\gamma}\rangle
\langle\psi_{N-1,\gamma}|\hat d_{i\Gamma\mu}
|\psi_{N\alpha}\rangle}{%
E_{N-1,\gamma}-E_{N\alpha} + \ek}=\\
-B^*_{k'i'\Gamma'}A_{ki\Gamma}\langle \psi_{N\beta}|
\hat d^{\dagger}_{i'\Gamma'\mu'}
\frac1{\hat{\mathcal H}_{ns}-E_{N\alpha} + \ek}
\hat d_{i\Gamma\mu}|\psi_{N\alpha}\rangle\,.
\end{multline}
Inserting these two expressions into Eq.~\eqref{eq:KHformulaSuppl} we
arrive at
\begin{multline}
I_-^{(\alpha)} = - \frac{2\pi e}{\hbar} \hskip-2em\sum_{\substack{%
\beta kk'ii'\Gamma\Gamma'\mu\mu'\\[.2em]
\ek > \eF+eV\text{ and } \ekk < \eF}}\hskip-2em
|A_{ki\Gamma}B_{k'i'\Gamma'}|^2
\biggl|
\langle \psi_{N\beta}|\hat d_{i\Gamma\mu}
\frac1{\hat{\mathcal H}_{ns}-E_{N\alpha}- \ekk}
\hat d^{\dagger}_{i'\Gamma'\mu'}
|\psi_{N\alpha}\rangle\\
-
\langle \psi_{N\beta}|
\hat d^{\dagger}_{i'\Gamma'\mu'}
\frac1{\hat{\mathcal H}_{ns}-E_{N\alpha} + \ek}
\hat d_{i\Gamma\mu}|\psi_{N\alpha}\rangle
\biggr|^2
\delta(\Delta_{\beta\alpha}+\ek-\ekk).
\end{multline}
The subsequent neglect of the $k$~dependence of the amplitudes $A$ and
$B$, the replacement of the momentum sums with integrals over the
single-particle energies, and setting $\eF=0$ yields
\begin{multline}
I_-^{(\alpha)} = - \frac{2\pi e}{\hbar} \sum_{\beta ii'\Gamma\Gamma'\mu\mu'}
|A_{i\Gamma}B_{i'\Gamma'}|^2
\int_{eV}^{\infty}\rmd\epsilon\, \rho^{t}_{i\Gamma}(\epsilon)
\int_{-\infty}^0 \rmd\epsilon'\, \rho^{s}_{i'\Gamma'}(\epsilon')\,
\delta(\Delta_{\beta\alpha}+\epsilon-\epsilon')\\
\times\biggl|
\langle \psi_{N\beta}|\hat d_{i\Gamma\mu}
\frac1{\hat{\mathcal H}_{ns}-E_{N\alpha}- \epsilon'}
\hat d^{\dagger}_{i'\Gamma'\mu'}
|\psi_{N\alpha}\rangle\\
-
\langle \psi_{N\beta}|
\hat d^{\dagger}_{i'\Gamma'\mu'}
\frac1{\hat{\mathcal H}_{ns}-E_{N\alpha} + \epsilon}
\hat d_{i\Gamma\mu}|\psi_{N\alpha}\rangle
\biggr|^2.
\end{multline}
Then, the differential conductance
$g_-^{(\alpha)}=\rmd I_-^{(\alpha)}/\rmd V$ becomes
\begin{multline}
g_-^{(\alpha)} = \frac{2\pi e^2}{\hbar} \sum_{\beta ii'\Gamma\Gamma'\mu\mu'}
|A_{i\Gamma}B_{i'\Gamma'}|^2 \rho^t_{i\Gamma}(eV)
\int_{-\infty}^0 \rmd\epsilon'\, \rho^s_{i'\Gamma'}(\epsilon')\,
\delta(\Delta_{\beta\alpha}+eV-\epsilon')\\
\times\biggl|
\langle \psi_{N\beta}|\hat d_{i\Gamma\mu}
\frac1{\hat{\mathcal H}_{ns}-E_{N\alpha}- \epsilon'}
\hat d^{\dagger}_{i'\Gamma'\mu'}
|\psi_{N\alpha}\rangle\\
-
\langle \psi_{N\beta}|
\hat d^{\dagger}_{i'\Gamma'\mu'}
\frac1{\hat{\mathcal H}_{ns}-E_{N\alpha} + eV}
\hat d_{i\Gamma\mu}|\psi_{N\alpha}\rangle
\biggr|^2.
\end{multline}
The remaining integral contributes only if
$\Delta_{\beta\alpha}+eV<0$. With this observation we finally come to
\begin{multline}
g_-^{(\alpha)} = \frac{2\pi e^2}{\hbar}
\hskipBSum
\sum_{\substack{\beta\\[.2em] \Delta_{\beta\alpha}<-eV}}\hskipBSum
\sum_{ii'\Gamma\Gamma'\mu\mu'}
|A_{i\Gamma}B_{i'\Gamma'}|^2
\rho^t_{i\Gamma}(eV) \rho^s_{i'\Gamma'}(eV+\Delta_{\beta\alpha})
\\
\times
\biggl|
\langle \psi_{N\beta}|\hat d_{i\Gamma\mu}
\frac1{\hat{\mathcal H}_{ns}-E_{N\beta}- eV}
\hat d^{\dagger}_{i'\Gamma'\mu'}
|\psi_{N\alpha}\rangle\\
-
\langle \psi_{N\beta}|
\hat d^{\dagger}_{i'\Gamma'\mu'}
\frac1{\hat{\mathcal H}_{ns}-E_{N\alpha} + eV}
\hat d_{i\Gamma\mu}|\psi_{N\alpha}\rangle
\biggr|^2.
\end{multline}
In terms of the transition operator $\hat{\mathcal O}$ defined in
Eq.~\eqref{eq:transOpSuppl}, the expression becomes
\begin{multline}
\label{eq:gMinusFinal}
g_-^{(\alpha)} = \frac{2\pi e^2}{\hbar}
\hskipBSum
\sum_{\substack{\beta\\[.2em] \Delta_{\beta\alpha}<-eV}}\hskipBSum
\sum_{ii'\Gamma\Gamma'\mu\mu'}
|A_{i\Gamma}B_{i'\Gamma'}|^2
\rho^t_{i\Gamma}(eV) \rho^s_{i'\Gamma'}(eV+\Delta_{\beta\alpha})\\[-1em]
\times
\bigl|
\langle \psi_{N\beta}|
\hat{\mathcal O}^{\beta\alpha}_{i'\Gamma'\mu'\,\, i\Gamma\mu}
|\psi_{N\alpha}\rangle
\bigr|^2.
\end{multline}



\subsection{Relation between the negative and positive currents}

Our final equations for the differential conductances,
Eqs.~\eqref{eq:gPlusFinal} and~\eqref{eq:gMinusFinal}, are very
similar but not exactly the same. We can observe that the matrix
element of the transition operator as it appears in
Eq.~\eqref{eq:gPlusFinal} can be transformed as
\begin{multline}
\bigl|\langle \psi_{N\beta}|
\hat{\mathcal O}^{\alpha\beta}_{i\Gamma\mu\,\, i'\Gamma'\mu'}
|\psi_{N\alpha}\rangle\bigr|^2
=
\bigl|\langle \psi_{N\alpha}|\bigl[
\hat{\mathcal O}^{\alpha\beta}_{i\Gamma\mu\,\, i'\Gamma'\mu'}
\bigr]^\dagger
|\psi_{N\beta}\rangle\bigr|^2
=\\
\bigl|\langle \psi_{N\alpha}|
\hat{\mathcal O}^{\alpha\beta}_{i'\Gamma'\mu'\,\, i\Gamma\mu}
|\psi_{N\beta}\rangle\bigr|^2,
\end{multline}
which is the corresponding expression from Eq.~\eqref{eq:gMinusFinal}
with $\alpha$ and $\beta$ interchanged. If we denote the whole summand
of Eq.~\eqref{eq:gPlusFinal} as
\begin{equation*}
\tag{\ref*{eq:partial_conductance}} 
{\mathcal G}_{\alpha\beta}
=
\frac{2\pi e^2}{\hbar}
\hskip-.5em\sum_{ii'\Gamma\Gamma'\mu\mu'}\hskip-.5em
|A_{i\Gamma}B_{i'\Gamma'}|^2
\rho_{i\Gamma}^t(eV)\rho_{i'\Gamma'}^s(eV-\Delta_{\beta\alpha})
\bigl|\langle \psi_{N\beta}|
\hat{\mathcal O}^{\alpha\beta}_{i\Gamma\mu\,\, i'\Gamma'\mu'}
|\psi_{N\alpha}\rangle\bigr|^2,
\end{equation*}
then the expressions for the differential conductances become
\begin{equation}
g_+^{(\alpha)} =
\sideset{}{_\beta}\sum_{\Delta_{\beta\alpha}<eV}
{\mathcal G}_{\alpha\beta}
\qquad\text{and}\qquad
g_-^{(\alpha)} =
\sideset{}{_\beta}\sum_{\Delta_{\beta\alpha}<-eV}
{\mathcal G}_{\beta\alpha}\,.
\end{equation}



\section{Fe adatom on Cu(100) probed by Nc-terminated tip}

\subsection{DFT calculations of Fe adatom on Cu(100) surface}

\begin{figure}
\centering
\includegraphics[width=0.65\linewidth]{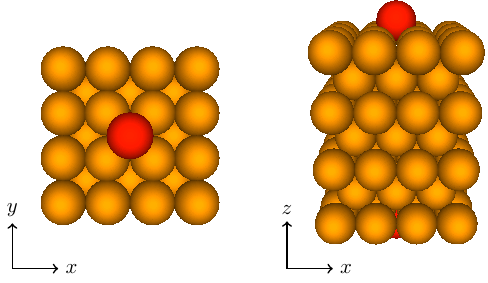}
\caption{\label{fig:DFTslab}Top (left) and side view (right) of the
Fe/Cu(100) unit cell as used in the DFT calculations. The iron atom is
absorbed on both sides of the slab to maintain as high symmetry as
possible (including inversion). The figure was made with VESTA
\cite{vesta}.}
\end{figure}

We have employed the density-functional theory (DFT) to estimate the
electronic structure of the iron adatom on the Cu(100) surface. Our
calculations were done with the
\textsc{wien2k} package \cite{wien2k} that implements the linearized
augmented plane-wave method and its extensions, and combines a
scalar-relativistic description with spin-orbit coupling. We used the
spin-polarized generalized-gradient approximation (PBE-GGA) for the
exchange-correlation functional \cite{perdew1996}.

We modeled the surface as a slab consisting of seven Cu layers. The
horizontal dimensions of the unit cell were
$7.668\text{~\AA}\times7.668\text{~\AA}$, corresponding to the
experimental lattice constant of fcc Cu $a=3.615$~\AA. The periodic
repetitions of the slab in the vertical direction were separated by
10.58~\AA\ ($20\, a_\text{B}$) of vacuum. The Fe adatom was placed in
the hollow position ($C_{4v}$~symmetry) on both sides of the slab such
that the unit cell was symmetric with respect to the inversion
(Fig.~\ref{fig:DFTslab}). The orientation of the coordinate system is
such that the in-plane 3d orbitals centered at the adatom behave as
follows: $xy$
points to the nearest neighbors and $x^2-y^2$ points to the voids
between the nearest neighbors. All internal coordinates in the
simulation cell were relaxed until the forces acting on the individual
atoms became smaller than 1~mRy/$a_\text{B}$. During the relaxation,
the spin-orbit coupling was switched off. The optimal adatom distance
above the surface was found to be~1.56~\AA.

The parameters determining the accuracy of the DFT calculations were:
the radii of the muffin-tin spheres were
$R_\text{MT}=2.07\,a_\text{B}$, the basis-set cutoff $K_\text{max}$ was
defined with $R_\text{MT}\times K_\text{max}=8.0$, and the Brillouin
zone was sampled on a $k$-point mesh $10\times 10\times 10$. The
default \textsc{wien2k} basis set with local orbitals for semicore
states Fe~3p and Cu~3p was used.

\begin{figure}
\centering
\includegraphics[width=\linewidth]{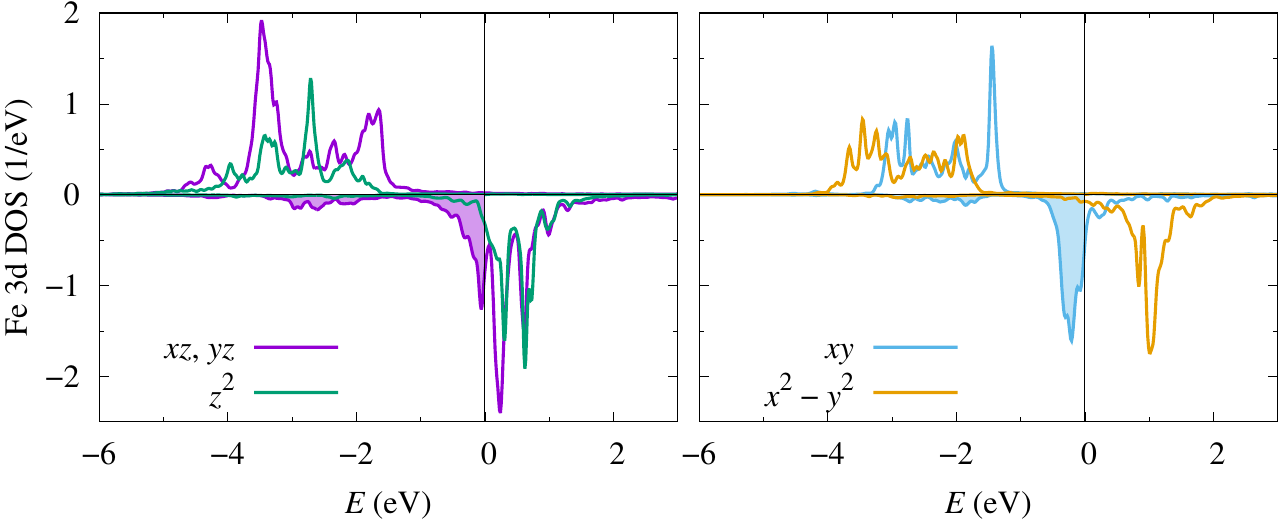}
\caption{\label{fig:DFT_DOS}Orbital-resolved density of states in the
  3d shell of the Fe adatom (a calculation without spin-orbit
  coupling). Majority spin is shown as positive values, minority spin
  as negative values.}
\end{figure}

The calculations give the spin moment at the Fe adatom as~$1.5$
(determined by integration of the spin density in the Fe muffin-tin
sphere, which may slightly underestimate the spin moment) or as
approximately~$1.6$ (determined as a half of the spin moment of the
whole unit cell, which counts also the small spin moments induced at
neighboring Cu atoms), which agrees with previous DFT calculations
\cite{stepanyuk2003,verlhac2019}. The density of states projected on
the adatom 3d orbitals is plotted Fig.~\ref{fig:DFT_DOS}, indicating
that the three unpaired electrons reside in the $z^2$ and $x^2-y^2$
orbitals, and in one of the doubly degenerate $xz$ and $yz$
orbitals.

After switching on the spin-orbit coupling, the magnetic
anisotropy energy can be estimated as a difference between the total
energies of the states with magnetic moments oriented in plane, [100]
direction, and out of plane, [001] direction. This way we obtain
$\Delta_\text{MAE}=E_{[100]}-E_{[001]}\approx 0.6$~meV/Fe, the sign of
which is consistent with experiments (the spin is out of plane in the
ground state) but the magnitude is an order of magnitude too small (if
our understanding of the inelastic tunneling spectra, detailed
in End Matter, is correct). We do not consider the underestimated
magnitude of $\Delta_\text{MAE}$ as a critical
contradiction, however, since DFT does not necessarily have the precision to
determine such small energy differences accurately, and we did not
check the convergence of $\Delta_\text{MAE}$ with respect to all
parameters, namely with respect to the size of the supercell, which is
computationally very costly and out of scope of the present study. We
only verified that $\Delta_\text{MAE}$ stays essentially the same when
the $k$-point mesh is increased to $15\times15\times15$.



\subsection{Spin model for coupled Ni and Fe spins}

\def\sp{\hskip.28em} 
\def\mix{30}         
\begin{sidewaystable}[htbp] 
\caption{\label{tbl:EneNcFe}
  Eigenstate energies $E$ and the gaps $\Delta=E-E_0$ for the spin model,
  Eq.~\eqref{eq:spin_modelSuppl}, expanded in powers of the exchange
  $J$ with only terms up to $J^2$ retained (the energy of state 2 is exact
  as shown). The energy increases from bottom to top when
  $|\DFe|\gg\DNc$, the color coding corresponds to
  Figs.~\ref{fig:spin_electron_model} and
  \ref{fig:spin_electron_model_diff_D}--\ref{fig:iets_spectra_diff_D_onlyNi}
  in the main text. The $z$~projection of the total spin $M_\text{tot}$
  and of the spins at the Ni and Fe atoms, $M_\text{Ni}$ and
  $M_\text{Fe}$, are listed as well; $M_\text{tot}$ is a
  conserving quantity (quantum number) since the model is axially
  symmetric around $z$, whereas $M_\text{Ni}$ and $M_\text{Fe}$ have
  the values as shown only at $J=0$.}
\small
\renewcommand{\arraystretch}{2.3}
\centering
\begin{tabular}{c|c|c|c|c|c}
\hline\hline
state & \sp$M_\text{Ni}$\sp{} & \sp$M_\text{Fe}$\sp{} &
\sp$M_\text{tot}$\sp{} & \sp$E$\sp{} & \sp$\Delta$ \\[.5em]
\hline
\rowcolor{colorE!\mix}
5 & $\pm 1$ & $\displaystyle\pm\frac12$ & $\displaystyle\pm\frac32$ &
  $\displaystyle \frac14 \DFe+\DNc +\frac12 J +
                                      \frac{3}{2\DNc-4\DFe}J^2$     &
  $\displaystyle -2\DFe+\DNc +\frac12 J +
                                      \frac{3}{\DNc-2\DFe}J^2$
\\[.5em]
\hline
\rowcolor{colorD!\mix}
4 & $\pm 1$ & $\displaystyle\pm\frac12$ & $\displaystyle\pm\frac12$ &
  $\displaystyle \frac14 \DFe+\DNc -\frac12 J + \frac{2}{\DNc}J^2$  &
  $\displaystyle -2\DFe+\DNc -\frac12 J + \biggl(\frac{2}{\DNc}
  + \frac{3}{2\DNc-4\DFe}\biggr)J^2$
\\[.5em]
\hline
\rowcolor{colorC!\mix}
3 &   0     & $\displaystyle\pm\frac12$ & $\displaystyle\pm\frac12$ &
  $\displaystyle
  \frac14 \DFe - \biggl(\frac{2}{\DNc}-\frac{3}{2\DNc+4\DFe}\biggr)J^2$&
  $\displaystyle
  -2\DFe - \biggl(\frac{2}{\DNc}
  + \frac{6\DFe}{(\DNc+2\DFe)(\DNc-2\DFe)}\biggr)J^2$
\\[.5em]
\hline
\rowcolor{colorB!\mix}
2 & $\pm 1$ & $\displaystyle\pm\frac32$ & $\displaystyle\pm\frac52$ &
  $\displaystyle \frac94 \DFe+\DNc +\frac32 J$ &
  $\displaystyle \DNc +\frac32 J
  + \frac{3}{2\DNc-4\DFe}J^2$
\\[.5em]
\hline
\rowcolor{colorA!\mix}
1 & $\pm 1$ & $\displaystyle\pm\frac32$ & $\displaystyle\pm\frac12$ &
  $\displaystyle \frac94 \DFe+\DNc -\frac32 J + \frac{3}{2\DNc+4\DFe}J^2$ &
$\displaystyle \DNc -\frac32 J + \frac{3\DNc}{(\DNc+2\DFe)(\DNc-2\DFe)}J^2$
\\[.5em]
\hline
\rowcolor{colorF!\mix}
0 &   0     & $\displaystyle\pm\frac32$ & $\displaystyle\pm\frac32$ &
  $\displaystyle \frac94 \DFe - \frac{3}{2\DNc-4\DFe}J^2$ &
  0
\\[.5em]
\hline\hline
\end{tabular}
\end{sidewaystable}

The anisotropic two-spin model for the coupled Ni and Fe spins,
\begin{equation*}
\tag{\ref*{eq:spin_model}} 
\label{eq:spin_modelSuppl}
\hat{\mathcal H}_\text{spin} = \DNc \hat S^2_{z,\text{Ni}}
 + \DFe \hat S^2_{z,\text{Fe}}
 - J \hat{\mathbf S}_\text{Ni}\cdot\hat{\mathbf S}_\text{Fe}\,,
\end{equation*}
allows for an analytic solution, which we used when plotting
Figs.~\ref{fig:spin_electron_model}
and~\ref{fig:spin_electron_model_diff_D}. The Hamiltonian
$\hat{\mathcal H}_\text{spin}$ commutes with the $z$~component of the
total spin,
$\hat S_{z,\text{tot}}=\hat S_{z,\text{Ni}} + \hat S_{z,\text{Fe}}$,
and hence the expectation value $M_\text{tot}$ of $\hat S_{z,\text{tot}}$
is a conserved quantity and the eigenstates with $\pm M_\text{tot}$
are degenerate. The half of the Hilbert space with positive
$M_\text{tot}$ decomposes into three subspaces: one-dimensional
$M_\text{tot}=5/2$, two-dimensional $M_\text{tot}=3/2$ and
three-dimensional $M_\text{tot}=1/2$. The eigenenergy for the
$M_\text{tot}=5/2$ state is
\begin{equation}
E_2=\frac94 \DFe+\DNc +\frac32 J\,,
\end{equation}
the eigenenergies in the $M_\text{tot}=3/2$ subspace are
\begin{equation}
E_{0,5}=\frac{5\DFe}{4} + \frac{\DNc}{2} + \frac{J}{4}
  \mp \frac{\DNc - 2\DFe}{2}\sqrt{1 + \frac{J}{\DNc
      -2\DFe} + \frac{25}{4}\biggl(\frac{J}{\DNc - 2\DFe}\biggr)^2}\,,
\end{equation}
and the eigenenergies in the $M_\text{tot}=1/2$ subspace are roots of
a cubic equation that do not have a compact enough form to be written
here explicitly. Instead, we list all eigenstate energies expanded in
powers of the exchange $J$ up to the second order in
Table~\ref{tbl:EneNcFe}.

If the exchange term
$J \hat{\mathbf S}_\text{Ni}\cdot\hat{\mathbf S}_\text{Fe}$ is
simplified to the Ising form
$J \hat{S}_{z,\text{Ni}}\,\hat{S}_{z,\text{Fe}}$, the Hamiltonian
$\hat{\mathcal H}_\text{spin}$ commutes with the $z$ projections of
the individual spins, $\hat{S}_{z,\text{Ni}}$ and
$\hat{S}_{z,\text{Fe}}$. This makes the eigenvalue problem trivial and
the exact eigenvalues are given by the expressions from
Table~\ref{tbl:EneNcFe} with the $J^2$ terms dropped
\cite{verlhac2019}.

%
%



\section{On the selection rules for the orbital moment}
\label{sec:orbital_exchange}

\begin{table}
\caption{\label{tbl:spinless_dshell}%
  The eigenstates and eigenenergies of a d shell hosting
  spinless electrons. Here $N$ is the filling
  of the shell, $|LM\rangle$ are states with the total orbital moment~$L$
  and its projection $M$ on the quantization axis. The other
  quantities are defined in the text. For the singly occupied shell to
  be the ground state, $\epsilon$ must be negative and $U$ larger than
  $|\epsilon|$.}
\centering
\def\arraystretch{1.4}
\begin{tabular}{cc|cc|cl}
\hline\hline
\multicolumn{2}{c|}{$N=0$} & 
\multicolumn{2}{c|}{$N=1$} &
\multicolumn{2}{c}{$N=2$}\\
\hline
\multirow{2}{*}{$|0\rangle$} & \multirow{2}{*}{0} &
\multirow{2}{*}{$|2,M\rangle$} & \multirow{2}{*}{$\epsilon$} &
$|3,M\rangle$ & $2\epsilon+U$ \\
 & & & & $|1,M\rangle$ & $2\epsilon+U+J_\text{H}$ \\
\hline\hline
\end{tabular}%

\end{table}

To analytically illustrate the selection rule for the total angular
moment, deduced numerically in the main text of the article, we
consider one
spinless electron in a d~shell (orbital momentum quantum number
$\ell=2$). In the absence of spin, the shell has only the orbital
moment. The tunneling proceeds via states with two electrons (orange
terms in the following formula) and via an empty shell (the blue
term). The transition operator defined in Eq.~\eqref{eq:transOpSuppl}
can be transformed to
\begin{multline}
\label{eq:transOspinless}
\hat{\mathcal O}_{mm'}^{\alpha\beta}
=
\biggl(
  \colorbox{blue_bg}{$\displaystyle \frac1{\epsilon}$}
-
  \colorbox{orange_bg}{$\displaystyle \frac{1}{\epsilon+U}$}
\biggr)
\hat d_m^\dagger\hat d_{m'}
+\colorbox{orange_bg}{$\displaystyle \frac1{\epsilon+U}\,\delta_{mm'}$}
\\
+ \colorbox{orange_bg}{$\displaystyle%
2\biggl(\frac1{\epsilon+U+J_\text{H}}-\frac{1}{\epsilon+U}\biggr)
\sum_{Mnn'}
\langle2n',2m'|1M\rangle
\langle1M|2n,2m\rangle
\hat d_{n'}^\dagger\hat d_n%
$}
,
\end{multline}
where $\epsilon<0$ is the single-particle energy of the
atomic level (the energy of the shell filled with one electron),
$U>|\epsilon|$ is the Coulomb repulsion, $J_\text{H}$ is the Hund
exchange in the
shell, and $\langle l_1m_1,l_2m_2|LM\rangle$ are Clebsch--Gordan
coefficients. The states of the d~shell relevant for the transition
operator are listed in Table~\ref{tbl:spinless_dshell}. The voltage
bias~$V$ was assumed to be much smaller than $U$
or $|\epsilon|$ and was neglected. The form of
Eq.~\eqref{eq:transOspinless} shows that
transitions from any $m'$ to any $m$ are possible, that is, $\Delta
m\leq 2\ell$. The same formula (for $J_{\rm H}=0$) was discussed in the
context of Ce impurities in \cite{coqblin1969}.



\section[Comparison to the theory of Delgado and Fern\'andez-Rossier]{%
  Comparison to Delgado and Fern\'andez-Rossier
  \cite{delgado2011}}

Our derivation of the differential conductance described in
Sec.~\ref{sec:dIdV_derivation} is based on assumptions that are very
similar to those of Delgado and Fern\'andez-Rossier in
\cite{delgado2011}, except that their theory is more general, since it
is applicable also at non-zero temperatures and is not limited to
small tunneling currents. After a closer inspection, however, we found that
their theory does not reduce to ours in the limit of zero temperature
and small currents. The origin of the discrepancies is a slightly
different operator describing the tunneling between the nanosystem and
the electrodes. The tunneling operator employed in \cite{delgado2011}
has the form
\begin{equation}
\hat{\mathcal{V}} = \sum_{kim\sigma} \Bigl(
A_{kim\sigma} \hat{a}^\dagger_{k\sigma}\hat{d}_{im\sigma}
+ A_{kim\sigma}^* \hat{d}_{im\sigma}^\dagger \hat{a}_{k\sigma}
+ B_{kim\sigma} \hat{b}^{\dagger}_{k\sigma}  \hat{d}_{im\sigma}
+ B_{kim\sigma}^* \hat{d}_{im\sigma}^\dagger \hat{b}_{k\sigma}
\Bigr)\,,
\label{eq:couplingDFR}
\end{equation}
which differs from our Eq.~\eqref{eq:couplingSuppl} by coupling the magnetic
centers directly to the Bloch waves in the electrodes instead of
the symmetry-conserving waves.

We now retrace the main steps of our derivation
(Sec.~\ref{sec:dIdV_derivation}) for the tunneling operator
$\hat{\mathcal{V}}$ given by
Eq.~\eqref{eq:couplingDFR}. The sums over the intermediate states
$|v\rangle$ slightly change, we explicitly show only the analogon of
Eq.~\eqref{eq:sum1IntermStatesPos},
\begin{multline}
\sideset{}{^{(1)}}\sum_v
\frac{\langle f_+|\hat{\mathcal V}|v\rangle
\langle v|\hat{\mathcal V}|\chi_{N\alpha}\rangle}{%
\mathcal E_{v}-\mathcal E_{N\alpha}}=\\
\sum_\gamma \frac{\langle \chi_{N\beta}|
\hat a^\dagger_{k\sigma}\hat b_{k'\sigma'}\hat{\mathcal V}
\hat a_{k\sigma}|\chi_{N+1,\gamma}\rangle
\langle \chi_{N+1,\gamma}|\hat a_{k\sigma}^\dagger
\hat{\mathcal V}|\chi_{N\alpha}\rangle}{%
E_{N+1,\gamma}-E_{N\alpha} - \ek}=\\
%
%
\sum_\gamma \frac{%
\langle \psi_{N\beta}|
\bigl(\sum_{i'm'}B_{k'i'm'\sigma'}\hat d_{i'm'\sigma'}\bigr)
|\psi_{N+1,\gamma}\rangle\langle \psi_{N+1,\gamma}|
\bigl(\sum_{im}A_{kim\sigma}^*\hat d^{\dagger}_{im\sigma}\bigr)
|\psi_{N\alpha}\rangle}{%
E_{N+1,\gamma}-E_{N\alpha} - \ek}=\\
\langle \psi_{N\beta}|
\underbrace{%
\Bigl(\sum_{i'm'}B_{k'i'm'\sigma'}\hat d_{i'm'\sigma'}\Bigr)%
}_{\displaystyle\hat B_{k'\sigma'}}
\frac1{\hat{\mathcal H}_{ns}-E_{N\alpha} - \ek}
\underbrace{%
\Bigl(\sum_{im}A_{kim\sigma}^*\hat d^{\dagger}_{im\sigma}\Bigr)%
}_{\displaystyle\hat A_{k\sigma}^\dagger}
|\psi_{N\alpha}\rangle\,.
\end{multline}
Because the tip and surface states coupled to the nanosystem do not
depend on indexes $i$ and $m$ now, the sum over these orbital degrees
of freedom originating from the tunneling operator $\hat{\mathcal{V}}$
survive inside the matrix element $\langle
\psi_{N\beta}|\bullet|\psi_{N\alpha}\rangle$.

The formula for the positive current, Eq.~\eqref{eq:posCurrentKSum},
changes to
\begin{multline}
I_+^{(\alpha)} = \frac{2\pi e}{\hbar}\hskip-2em\sum_{\substack{%
\beta kk'\sigma\sigma'\\[.2em]
\ek < \eF+eV\text{ and } \ekk > \eF}}\hskip-2em
\biggl|
\langle \psi_{N\beta}|
\hat B_{k'\sigma'}
\frac1{\hat{\mathcal H}_{ns}-E_{N\alpha} - \ek}
\hat A_{k\sigma}^\dagger
|\psi_{N\alpha}\rangle\\
-
\langle \psi_{N\beta}|
\hat A_{k\sigma}^\dagger
\frac1{\hat{\mathcal H}_{ns}-E_{N\alpha} + \ekk}
\hat B_{k'\sigma'}
|\psi_{N\alpha}\rangle
\biggr|^2
\delta(\Delta_{\beta\alpha}-\ek+\ekk)\,,
\end{multline}
where the outer sum, that is, the sum over the final states
$|f_+\rangle$, now does not involve the orbital degrees of freedom $i$
and $m$, since the final state does not depend on them. When the
$k$ dependence of the tunneling amplitudes $A$ and $B$ is neglected,
which we do in Sec.~\ref{sec:dIdV_derivation} and Delgado and
Fern\'andez-Rossier do in \cite{delgado2011}, the differential
conductance takes the form
\begin{equation}
\label{eq:gPlusFinalDFR}
g_+^{(\alpha)} = \frac{2\pi e^2}{\hbar} \hskipBSum
\sum_{\substack{\beta\\[.2em]\Delta_{\beta\alpha}<eV}}\hskipBSum
\sum_{\sigma\sigma'}
\rho^t(eV)\rho^s(eV-\Delta_{\beta\alpha})
\bigl|\langle \psi_{N\beta}|
\hat{\mathcal O}^{\alpha\beta}_{\sigma\sigma'}
|\psi_{N\alpha}\rangle\bigr|^2,
\end{equation}
where the transition operator is
\begin{multline}
\label{eq:transOpDFR}
\hat{\mathcal O}^{\alpha\beta}_{\sigma\sigma'}
=
\colorbox{orange_bg}{$\displaystyle
\Bigl(\sum_{i'm'}B_{i'm'\sigma'}\hat d_{i'm'\sigma'}\Bigr)
\frac1{\hat{\mathcal H}_{ns} - E_{N\alpha} - eV}
\Bigl(\sum_{im}A_{im\sigma}^*\hat d^{\dagger}_{im\sigma}\Bigr)%
$}
\\
-
\colorbox{blue_bg}{$\displaystyle
\Bigl(\sum_{im}A_{im\sigma}^*\hat d^{\dagger}_{im\sigma}\Bigr)
\frac1{\hat{\mathcal H}_{ns} - E_{N\beta} + eV}
\Bigl(\sum_{i'm'}B_{i'm'\sigma'}\hat d_{i'm'\sigma'}\Bigr)%
$}
\,.
\end{multline}
The last two equations have the same structure as the equations
listed in Appendix~B of \cite{delgado2011}. Note that $i$ denotes
\emph{all} orbital degrees of freedom in \cite{delgado2011}, not just
the site.

\begin{figure}
\includegraphics[width=\linewidth]{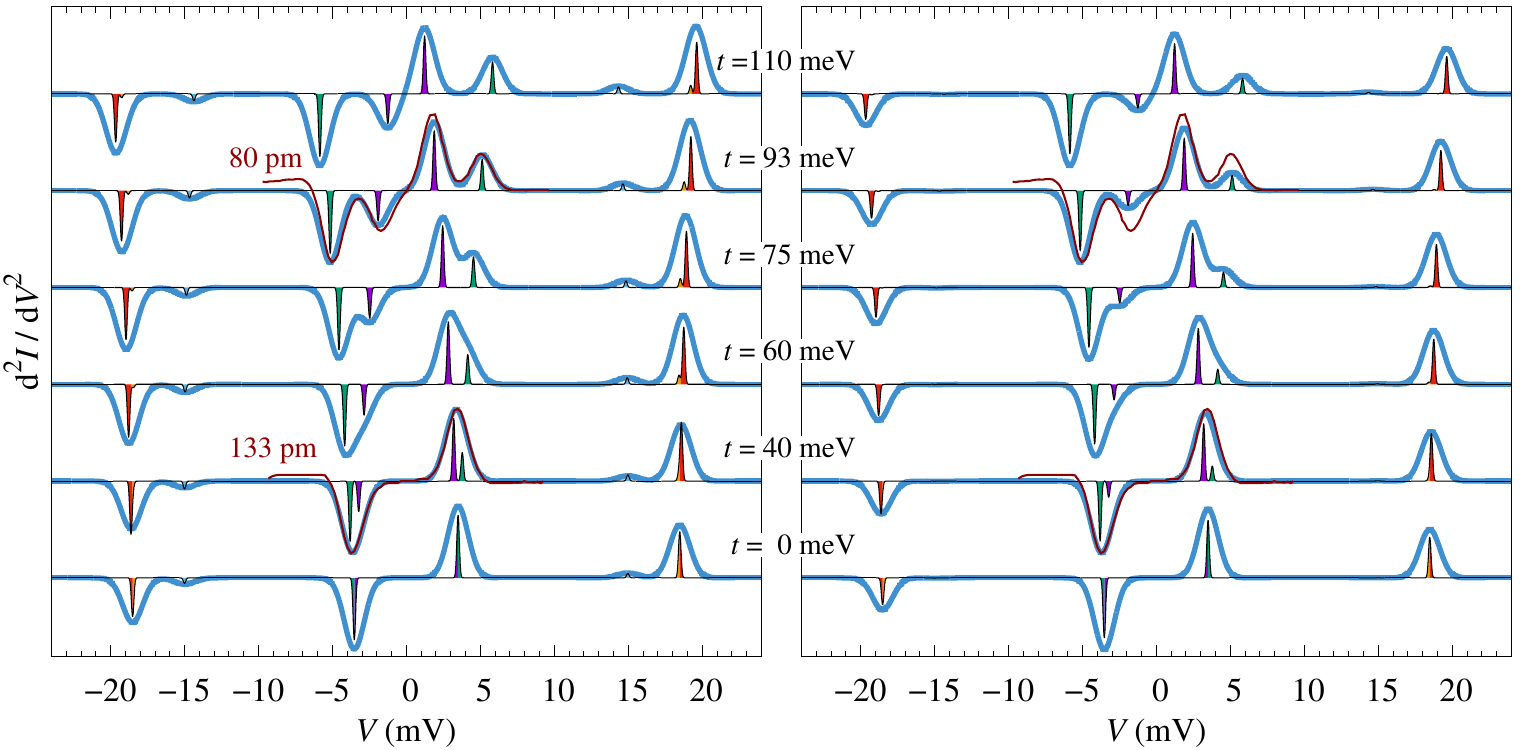}
\caption{\label{fig:FeCu100_DFR}Comparison of the $\rmd^2I/\rmd V^2$
  spectra calculated with our tunneling Hamiltonian,
  Eq.~\eqref{eq:couplingSuppl}, (left) and with the tunneling Hamiltonian of
  Delgado and   Fern\'andez-Rossier \cite{delgado2011},
  Eq.~\eqref{eq:couplingDFR}, (right). The system is Fe adatom on the
  Cu(100) surface probed by a nickelocene terminated tip, the parameters of
  the atomic shells are listed in rows A and~C in
  Table~\ref{tbl:electron_parameters} (the same
  $\hat{\mathcal H}_{ns}$ as in Fig.~\ref{fig:iets_spectra_diff_D}).}
\end{figure}

We have implemented these equations and applied them to the same cases
that are discussed in the main text. Figure~\ref{fig:FeCu100_DFR}
shows the calculated $\rmd^2I/\rmd V^2$ spectra corresponding to Fe a
adatom on the Cu(100) surface probed by a nickelocene-decorated tip. The
differences between the tunneling Hamiltonian from \cite{delgado2011},
Eq.~\eqref{eq:couplingDFR}, and our tunneling Hamiltonian,
Eq.~\eqref{eq:couplingSuppl}, are not dramatic, but they are certainly
visible -- and the match to the experimental $\rmd^2I/\rmd V^2$
spectra \cite{verlhac2019} is distinctly worse with the Hamiltonian
employed in \cite{delgado2011}.

In the case of a partially filled atomic shell placed in a homogeneous
magnetic field~$\vec B$, the nanosystem Hamiltonian is just
\begin{equation}
\hat{\mathcal H}_{ns}=\sum_{mm'\sigma\sigma'}
\bigl[\zeta(\vec l\cdot\vec s)
-\mu_B (\vec l+2\vec s)\cdot \vec B\bigr]^{mm'}_{\sigma\sigma'}
\hat d_{m\sigma}^\dagger \hat d_{m'\sigma'}
+\hat U\,.
\end{equation}
When the nanosystem--electrode tunneling amplitudes are taken in the
simple form $A_{m\sigma}=A$ and $B_{m\sigma}=B$, the system should be
entirely isotropic with the only direction-dependent term being the
interaction with the magnetic field. Yet, the tunneling Hamiltonian from
\cite{delgado2011} yields \emph{different} $\rmd^2I/\rmd V^2$ spectra for
different directions of the magnetic field (say along $z$ and along
$x$), which is clearly \emph{inconsistent} with the symmetry of the
system. Our tunneling Hamiltonian, Eq.~\eqref{eq:couplingSuppl}, on the other
hand, gives $\rmd^2I/\rmd V^2$ spectra independent of the direction of
the magnetic field. The different outcomes of the two tunneling
Hamiltonians can be traced to the property that the tunneling events
described by Eq.~\eqref{eq:couplingSuppl} conserve spin as well as orbital
moment, the tunneling described by Eq.~\eqref{eq:couplingDFR}
conserves only spin.

The spurious directional dependence can be analytically illustrated on
the d~shell hosting one spinless electron discussed in
Sec.~\ref{sec:orbital_exchange}. To keep things as simple as possible,
we consider the limit $U\to\infty$, which makes the two-electron
intermediate states inaccessible and the tunneling can proceed only
via the empty shell. Our transition operator,
Eq.~\eqref{eq:transOspinless}, then simplifies to
$\hat{\mathcal O}_{mm'}^{\alpha\beta}
=\epsilon^{-1}\hat d_m^\dagger\hat d_{m'}$,
and the differential conductance (positive current)
is given by
\begin{multline}
g_+^{(\alpha)} \sim
\frac1{\epsilon^2}\hskipBSum
\sum_{\substack{\beta\\[.2em]\Delta_{\beta\alpha}<eV}}\hskipBSum
\sum_{mm'}
\bigl|\langle\beta|\hat d_m^\dagger\hat d_{m'}|\alpha\rangle
\bigr|^2
\equiv
\frac1{\epsilon^2}\hskipBSum
\sum_{\substack{\beta\\[.2em]\Delta_{\beta\alpha}<eV}}\hskipBSum
\sum_{mm'}
\bigl|\langle\beta|\hat d_m^\dagger|0\rangle\langle0|
  \hat d_{m'}|\alpha\rangle
\bigr|^2
= \\
\frac1{\epsilon^2}\hskipBSum
\sum_{\substack{\beta\\[.2em]\Delta_{\beta\alpha}<eV}}
\underbrace{\Bigl(\sum_m
\bigl|\langle\beta|\hat d_m^\dagger|0\rangle\bigr|^2\Bigr)}_{=\,1}
\underbrace{\Bigl(\sum_{m'}
\bigl|\langle0|\hat d_{m'}|\alpha\rangle\bigr|^2\Bigr)}_{=\,1}
=
\hskipBSum
\sum_{\substack{\beta\\[.2em]\Delta_{\beta\alpha}<eV}}\hskipBSum
\frac1{\epsilon^2}\,,
\end{multline}
where operators $\hat d_m$ correspond to eigenstates of the $z$
component of the orbital moment~$\hat l_z$, and $|\alpha\rangle$ and
$|\beta\rangle$ are one-electron eigenstates of the shell Hamiltonian.
The sums over $m$ and $m'$ in the second line are just
normalization conditions of the states $|\alpha\rangle$ and
$|\beta\rangle$, since the states $\hat d_m$ form a basis of the shell. The
final expression for the differential conductance does not depend on
any details of the eigenstates, and hence it is invariant with respect
to the direction of the magnetic field applied to the shell.

The alternative transition operator, Eq.~\eqref{eq:transOpDFR}, has
the form
\begin{equation}
\hat{\mathcal O}^{\alpha\beta}=\frac{1}{\epsilon}
 \Bigl(\sum_{m}A^* \hat d^{\dagger}_{m}\Bigr)
 \Bigl(\sum_{m'}B\, \hat d_{m'}\Bigr)
\equiv
\frac{1}{\epsilon}
 \Bigl(\sum_{m}A^* \hat d^{\dagger}_{m}\Bigr)|0\rangle\langle0|
 \Bigl(\sum_{m'}B\, \hat d_{m'}\Bigr)\,,
\end{equation}
and the corresponding formula for the differential conductance,
Eq.~\eqref{eq:gPlusFinalDFR}, reads as
\begin{equation}
\label{eq:gPlusSpinless}
g_+^{(\alpha)} \sim
\frac1{\epsilon^2}\hskipBSum
\sum_{\substack{\beta\\[.2em]\Delta_{\beta\alpha}<eV}}\hskipBSum
\biggl|\langle\beta|
\Bigl(\sum_{m}\hat d^{\dagger}_{m}\Bigr)|0\rangle\langle0|
 \Bigl(\sum_{m'}\hat d_{m'}\Bigr)
|\alpha\rangle\biggr|^2\,.
\end{equation}
When the magnetic field is oriented along the $z$ direction, the
eigenstates of the shell are simply
$|\beta\rangle=\hat d_\beta^\dagger|0\rangle$ and the differential
conductance again becomes
\begin{equation}
\label{eq:gPlusBz}
g_+^{(\alpha)} \sim \hskipBSum
\sum_{\substack{\beta\\[.2em]\Delta_{\beta\alpha}<eV}}\hskipBSum
\frac1{\epsilon^2}\,,
\end{equation}
since only one term from each of the sums over $m$ and $m'$ contributes.
When the magnetic field is oriented along the $x$ direction, the
eigenstates of the shell can be obtained by rotating the $B_z$
eigenstates by $-\pi/2$ around the $y$~axis (this rotation transforms
the $z$~axis to the $x$~axis), namely
$|\tilde\beta\rangle=\exp(-i\pi\,\hat l_y/2)\,\hat d_\beta^\dagger|0\rangle$.
The differential conductance, Eq.~\eqref{eq:gPlusSpinless}, then
transforms to
\begin{multline}
g_+^{(\alpha)} \sim \frac1{\epsilon^2}\hskipBSum
\sum_{\substack{\beta\\[.2em]\Delta_{\beta\alpha}<eV}}\hskipBSum
\biggl|\sum_m\langle 0|\hat d_\beta\, \rme^{i\pi\,\hat l_y/2}
  \hat d_m^\dagger|0\rangle\biggr|^2
\biggl|\sum_{m'}\langle 0|\hat d_{m'}\,
  \rme^{-i\pi\,\hat l_y/2}\,\hat d_\alpha^\dagger|0\rangle\biggr|^2
=\\
\frac1{\epsilon^2}\hskipBSum
\sum_{\substack{\beta\\[.2em]\Delta_{\beta\alpha}<eV}}\hskipBSum
\biggl|\sum_m\langle\beta|\rme^{i\pi\,\hat l_y/2}
  |m\rangle\biggr|^2
\biggl|\sum_{m'}\langle m'|
  \rme^{-i\pi\,\hat l_y/2}|\alpha\rangle\biggr|^2.
\label{eq:gPlusBx}
\end{multline}
It involves modulus squared of sums of columns/rows of the unitary matrix
\begin{equation}
e^{-i\pi\,\hat l_y/2}=
\def\arraystretch{1.4}
\begin{pmatrix}
\frac{1}{4} & \frac{1}{2} & \sqrt{\frac{3}{8}} & \frac{1}{2} & \frac{1}{4} \\
 -\frac{1}{2} & -\frac{1}{2} & 0 & \frac{1}{2} & \frac{1}{2} \\
 \sqrt{\frac{3}{8}} & 0 & -\frac{1}{2} & 0 & \sqrt{\frac{3}{8}} \\
 -\frac{1}{2} & \frac{1}{2} & 0 & -\frac{1}{2} & \frac{1}{2} \\
 \frac{1}{4} & -\frac{1}{2} & \sqrt{\frac{3}{8}} & -\frac{1}{2} & \frac{1}{4}
\end{pmatrix}
\end{equation}
and is clearly different from Eq.~\eqref{eq:gPlusBz}. The sums of some rows
are zero and the corresponding terms entirely drop from the sum over
$\beta$ in Eq.~\eqref{eq:gPlusBx}.



\bibliography{IETS_HubI}